\documentclass[twocolumn, twocolappendix]{aastex631}
\usepackage{mathptmx}
\usepackage[T1]{fontenc}
\usepackage{newtxtext,newtxmath}
\usepackage{graphicx}	
\usepackage{amsmath}	
\usepackage{float}
\usepackage{comment,color}
\usepackage{pgfplotstable, booktabs, array}
\usepackage[ruled,vlined]{algorithm2e}
\usepackage{ulem}
\usepackage{footnote}
\usepackage{afterpage}
\usepackage{hyperref}

\pgfplotstableread[col sep=space]{kids_50lens_phot_cata.dat}\datatable
\newcommand{\Zhong}[1]{{\color{black}{#1}}}

\begin{document}
\title{Galaxy-Galaxy Strong Lensing with U-Net (GGSL-UNet). I. Extracting 2-Dimensional Information from Multi-Band Images in Ground and Space Observations}

\correspondingauthor{Nicola R. Napolitano, Ruibiao Luo}
\email{nicolarosario.napolitano@unina.it, luorb@pmo.ac.cn}

\author{Fucheng Zhong}
\affiliation{School of Physics and Astronomy, Sun Yat-sen University, Zhuhai Campus, 2 Daxue Road, Xiangzhou District, Zhuhai, 519082, P. R. China}

\author{Ruibiao Luo}
\affiliation{Purple Mountain Observatory, Chinese Academy of Sciences, Nanjing 210023, P. R. China}

\author{Nicola R. Napolitano}
\affiliation{Department of Physics E. Pancini, University Federico II, Via Cinthia 21, I-80126, Naples, Italy}
\affiliation{INAF – Osservatorio Astronomico di Capodimonte, Salita Moiariello 16, I-80131 Napoli, Italy}

\author{Crescenzo Tortora}
\affiliation{INAF – Osservatorio Astronomico di Capodimonte, Salita Moiariello 16, I-80131 Napoli, Italy}

\author{Rui Li}
\affiliation{Institude for Astrophysics, School of Physics, Zhengzhou University, Zhengzhou, 450001, China}

\author{Xincheng Zhu}
\affiliation{Department of Astronomy, Tsinghua University, Beijing 100084, China}

\author{Valerio Busillo}
\affiliation{Department of Physics E. Pancini, University Federico II, Via Cinthia 21, I-80126, Naples, Italy}
\affiliation{INAF – Osservatorio Astronomico di Capodimonte, Salita Moiariello 16, I-80131 Napoli, Italy}

\author{L. V. E. Koopmans}
\affiliation{Kapteyn Astronomical Institute, University of Groningen, P.O.Box 800, 9700AV Groningen, the Netherlands}

\author{Giuseppe Longo}
\affiliation{Department of Physics E. Pancini, University Federico II, Via Cinthia 21, I-80126, Naples, Italy}

\begin{abstract}
    We present a novel deep learning method to separately extract the two-dimensional flux information of the foreground galaxy (deflector) and background system (source) of Galaxy-Galaxy Strong Lensing events using U-Net (GGSL-Unet for short). In particular, the segmentation of the source image is found to enhance the performance of the lens modeling, especially for ground-based images.
    By combining mock lens foreground+background components with real sky survey noise to train the GGSL-Unet, we show it can correctly model the input image noise and extract the lens signal. However, the most important result of this work is that the GGSL-UNet can accurately reconstruct real ground-based lensing systems from the Kilo Degree Survey (KiDS) in one second.
    We also test the GGSL-UNet on space-based (HST) lenses from BELLS GALLERY, and obtain comparable accuracy 
    of standard lens modeling tools.
    Finally, we calculate the magnitudes from the reconstructed deflector and source images and use this to derive photometric redshifts (photo-$z$), with the photo-$z$ of the deflector well consistent with spectroscopic ones.
   This first work, demonstrates the great potential of the generative network for lens finding, image denoising, source segmentation, and decomposing and modeling of strong lensing systems.
    For the upcoming ground- and space-based surveys, the GGSL-UNet can provide high-quality images as well as geometry and redshift information for precise lens modeling, in combination with classical MCMC modeling for best accuracy in the galaxy-galaxy strong lensing analysis.
\end{abstract}

\keywords{galaxy-galaxy lens --- photometric --- deep learning --- image reconstruction.}

\section{Introduction}

Galaxy-galaxy lensing (GGL) has become an indispensable tool for studying the evolution of the universe, the nature of dark matter, and the structure and properties of galaxies \citep{Treu2010ARA&A..48...87T, Hoekstra2013SSRv..177...75H}. 

Gravitational lensing is a phenomenon involving foreground galaxies (named lenses or deflectors) and background far-away systems (sources), where the gravitational field of the former total mass component distorts images of these latter systems. Depending on the mass of the deflector and the alignment of the source and the foreground galaxy, the GGL can be either ``weak'' or ``strong.'' In the ``strong'' version of this phenomenon, the lensed image of the sources can exhibit different symmetries depending on the relative positions of the background and foreground galaxies along the line of sight and the intrinsic shape of the source itself: a double or quadruply image or even an ``Einstein cross'' (in case the four images form a perfect cross) in case the source is a compact system (e.g., a quasar, a galactic nucleus or even a supernova), a closed ring (aka ``Einstein ring'') for a perfect alignment, or a semi-arc \citep{1992grle.book.....S, 1998LRR.....1...12W} for a certain degree of alignment, in case of a more extended source (e.g., a galaxy).  
For larger misaligned, the GGL becomes ``weak,'' producing small deformations of the background sources, of the order of the thousandths of time of the original ellipticity, over large areas around the lens. 

The ``weak'' GGL provides a method for estimating the excess surface density and the halo profile with its statistics (e.g., \citealt{Clampitt2017MNRAS.465.4204C,luo2018}; \citealt{Luo2022A&A...668A..12L, Dvornik2023A&A...675A.189D, Chaurasiya2024MNRAS.527.5265C}), helping reconstruct the cluster mass distribution \citep{2001PhR...340..291B, 2005A&A...437...39B}, and constraining the cosmological parameters \citep{2009MNRAS.394..929C}. The ``strongly'' lensed and magnified background galaxies, instead, offer a direct way to constraint the Hubble constant through the time delay of source variability \citep{1964MNRAS.128..307R, 2003ApJ...599...70K} and investigate very high redshift galaxies during the reionization epoch of the universe (e.g., \citealt{2013ApJ...762...32C, Atek2015ApJ...800...18A, Kawamata2016ApJ...819..114K,2019ApJ...884...85C, 2023AJ....165...13W}). By incorporating the assumption of dark matter particle interactions on small scales, ``strong'' GGL can also provide an additional method for constraining different dark matter models (e.g., \citealt{2020MNRAS.491.6077G, 2021MNRAS.507.2432G, 2023PhRvD.107j3008G}). Obviously, the lensing phenomenon helps to map the fine details of the mass distribution of the foreground galaxy, such as dark matter substructures (e.g., \citealt{Vegetti2009MNRAS.392..945V, Vegetti2010MNRAS.408.1969V, Vegetti2014MNRAS.442.2017V}), naturally providing strong constraints on its total mass, and dark matter fraction \citep{Koopmans2006ApJ...649..599K, Gavazzi2007ApJ...667..176G, Vegetti2009MNRAS.400.1583V, 2010ApJ...724..511A, Treu2010ARA&A..48...87T, Tortora2010ApJ...721L...1T, Saha2024SSRv..220...12S}. 

Focusing on the strong lensing, with the advent of the III-Stage sky surveys (e.g., the Kilo Degree Survey, KiDS, \citealt{2013ExA....35...25D}; the DeCAM Legacy Survey, DeCALs, \citealt{2015AJ....150..150F,2016MNRAS.460.1270D}; the Hyper Suprime Camera Survey, HSC, \citealt{2018PASJ...70S...4A}) and the upcoming IV-Stage Surveys (e.g., Euclid \citealt{2011arXiv1110.3193L}; Vera Rubin Legacy Survey in Space and Time, LSST, \citealt{2019ApJ...873..111I}); the China Space Station Telescope Survey, CSST, \citealt{2011SSPMA..41.1441Z}), it has become mandatory to seek for the highest completeness in the lensing search to increase the number of lens targets and achieve statistical significance in scientific analysis, many automated strong lens detection tools have been developed (e.g., \citealt{
2006ApJ...638..703B, 2013ApJ...777...98S, 2014ApJ...785..144G, 2017MNRAS.472.1129P, 2018ApJ...856...68P, Petrillo2019a_MNRAS.482..807P,Petrillo2019b_MNRAS.484.3879P,Li2020ApJ...899...30L,2021ApJ...923...16L,Thuruthipilly2022arXiv221212915T,2023MNRAS.523.4188N,Thuruthipilly2024IAUS..381...28T}), including both classical methods and deep-learning-based approaches. Thousands of lens candidates have been found, though a small fraction of them have been modeled.
Modeling lenses often require images with a high enough Signal-to-Noise Ratio (SNR), such as those from the Hubble Space Telescope (HST) in the Sloan Lens ACS Survey (SLACS, \citealt{2006ApJ...638..703B,2008ApJ...682..964B, shu2015ApJ...803...71S,2017ApJ...851...48S}) or the Baryon Oscillation Spectroscopic Survey (BOSS) Emission-Line Lens Survey (BELLS) for the GALaxy-Ly$\alpha$ EmitteR sYstems (BELLS GALLERY, \citealt{Shu2016ApJ...833..264S}), as well as those from the HSC, such as the Survey of Gravitationally-Lensed Objects in HSC Imaging (SuGOHI, \citealt{2019A&A...630A..71S}). However, this process can be time-consuming and often requires empirical initial parameters. Furthermore, it is prone to degeneracies between the parameters of the foreground and background galaxy light, which can bias the mass inference. With the rapid development of generative neural networks, these networks can reconstruct images from given inputs, providing us with a novel way to denoise and model lensing objects.
For instance, if we can model the different lens components with high accuracy, one can use the background light only to increase the speed and accuracy of standard MCMC modeling. This will help us to estimate the morphology parameters of the arc \citep{2023MNRAS.522.5442G}, as well as the redshift and photometric parameters \citep{2011MNRAS.413.1753Z, 2021A&A...655A..81P}.
To provide these different lens components for the subsequent modeling task, we use U-Net \citep{2015arXiv150504597R}, a generative deep learning model, to reconstruct noise-reduced images of the foreground and background sources, respectively, from the lens image input.

Using U-Net, we can reduce model dependency in lens modeling, decrease the number of parameters required, model the lens quickly, and provide key information, such as magnitude and photometric redshift (photo-z), for lens modeling. Through the reconstructed image, we can calculate and compare the $\chi^2$ of different components transparently rather than directly obtaining lens parameters via a supervised deep learning model, which often appears as an unknown ``black box.'' By knowing the flux from different components, we can even trace the foreground total mass using the geometry of the source flux distribution and trace the stellar mass using the foreground flux.

To evaluate the effectiveness of the reconstructed images, $\chi^2$ will be calculated in the subsequent analysis. To demonstrate that the model trained on simulation data can be applied to real data, dozens of real lens images, including lens candidates from KiDS with corresponding spectroscopy measurements and BELLS GALLERY lenses from HST with modeling (see \S\ref{sec: real lens}), are used as test samples. We also compare the reconstructed results with those obtained using the previous method based on Markov-Chain-Monte-Carlo (MCMC).

The contents of this paper are as follows: \S \ref{sec: data} introduces the dataset, including both simulated and real data. \S \ref{sec: Method} covers the basic concept and technical details of reconstruction using U-Net. \S \ref{sec: Evaluation} presents the training results, the reconstructions on mock and real lenses, and a comparison with previous methods. In \S \ref{sec: Downstream task}, we provide an example of a downstream task: predicting photometric redshift using the reconstructed images as input. Finally, in \S \ref{sec: Conclusion}, some discussions and conclusions are provided.

\section{Data}
\label{sec: data}
The main aim of this work is to train a UNet to accurately separate the signal from the lens, from the source, and the noise of a strong GGL system. In this section, we discuss the best strategies we have decided to adopt to efficiently train our Galaxy-Galaxy Strong Lensing with U-Net (GGSL-UNet here after) tool and the data used for the network's training and testing phases. 

\subsection{Real or mock training samples?}
\label{sec:real_or_mock}
There are a few thousand strong GGL candidates cataloged by different observational programs, such as SLACS (mentioned above), Strong Lensing Legacy Survey (SL2S, \citealt{2013ApJ...777...98S}), HSC lens candidates \citep{2018PASJ...70S..29S, 2021A&A...653L...6C, 2022A&A...662A...4S}, DES \citep{2016ApJ...827...51N, 2017ApJS..232...15D, 2019ApJS..243...17J}, DECALS  \citep{2020ApJ...894...78H, 2021ApJ...909...27H, 2022ApJ...932..107S, 2022arXiv220602764S}, and KiDS \citep{2019MNRAS.484.3879P, Li2020ApJ...899...30L, 2021ApJ...923...16L}. The latest and largest catalogs have been more and more based on deep-learning techniques based on photometric data (for instance, \citealt{2017MNRAS.472.1129P, Li2020ApJ...899...30L}) or even spectroscopy data (see \citealt{2019MNRAS.482..313L}, and GaSNet \citealt{2022RAA....22f5014Z}).
Despite the large amount of high-quality candidates collected, these are insufficient to be directly used as the training set for DL models as they are far fewer than the requirements for DL model training. More importantly, it is impossible to define robust enough model parameters for them to be used as ``ground truth'' for network training. The parameters of modeled lenses are derived from certain codes, which might introduce systematic biases inherent to the code rather than totally reflecting the ground truth. These biases could be learned and inherited by the network trained on such data.

Using simulated lenses as training data is one way to solve this problem.
Combining mock foreground and background sources with real sky survey image noise cutouts can produce data with realistic noise. By designing the network to fit the mock and the real noise components, one can expect to reconstruct the different components of both mock and real lenses. The critical point is that if the network can fit the real noise component very well, even when a mock or not fully realistic component is used in the training data, we can expect the rest of the components to be the signal we want after noise fitting. We can produce the lens images within the specified observation conditions and cover as many major physical situations as possible. In the next subsection, we introduce the main observation datasets that used in this paper.

\pgfkeys{/pgf/number format/.cd,
fixed, zerofill,
precision=2,
set thousands separator={\,},
1000 sep in fractionals,
}
\begin{table*}
    \scalebox{0.68}{\hspace{-2cm}
    \pgfplotstabletypeset[
        multicolumn names,
        columns={KiDS_ID, RA, DEC, BPZ, BPZ_MIN, BPZ_MAX, paper, z, SURVEY, MAG_GAaP_u,MAG_GAaP_g,MAG_GAaP_r,MAG_GAaP_i},
        columns/KiDS_ID/.style={column name=ID, string type},
        columns/RA/.style={column name=RA},
        columns/DEC/.style={column name=DEC},
        columns/BPZ/.style={column name=$z_{BPZ}$, string type},
        columns/BPZ_MIN/.style={column name=$z_{BPZ}(MIN)$, string type},
        columns/BPZ_MAX/.style={column name=$z_{BPZ}(MAX)$, string type},
        columns/paper/.style={column name=$paper$, string type},
        columns/z/.style={column name=$z_{l,spec}$},
        columns/SURVEY/.style={column name=SURVEY, string type},
        columns/MAG_GAaP_u/.style={column name=$MAG_{GAaP}(u)$},
        columns/MAG_GAaP_g/.style={column name=$MAG_{GAaP}(g)$},
        columns/MAG_GAaP_r/.style={column name=$MAG_{GAaP}(r)$},
        columns/MAG_GAaP_i/.style={column name=$MAG_{GAaP}(i)$},
        every head row/.style={before row={\toprule\hline&{}&{}&{}&{}&{}&{}&{}&{}&{}&{}&{}&{}\\}, after row/.add={}{%
        \arraybackslash%
        {}&{}&{}&{}&{}&{}&{}&{}&{}&{}&{}&{}&{}\\
         \midrule\hline}
        },
        every last row/.style={after row=\hline\bottomrule},
        ]\datatable
    }
    \caption{Basic information for the 50 selected cases from KiDS-DR4. Here are the object ID and position in KiDS. It also shows the photo-z ($z_{BPZ}$) with lower ($z_{BPZ}(MIN)$) and upper ($z_{BPZ}(MAX)$) limits estimated by BPZ software based on 9-band photometric data ($ugri$ from KiDS, and $ZYJHK_S$ from VIKING). The $paper$ column indicates the publication source of the strong lens candidates: Li$+$20 \citep{Li2020ApJ...899...30L}, P$+$19 \citep{Petrillo2019a_MNRAS.482..807P, Petrillo2019b_MNRAS.484.3879P}, with `bonus' indicating follow-up lens candidates from P$+$19. The $z_{l,spec}$ shows the spectroscopic redshifts from SDSS (eBOSS, \citealt{Alam2021PhRvD.103h3533A}), BOSS \citep{Dawson2013AJ....145...10D}, GAMA \citep{Driver2022MNRAS.513..439D}, and MUSE \citep{2020ApJ...904L..31N}. The last four columns provide the optical ($ugri$) GAaP magnitudes.
    }
    \label{table:51_KiDS_lens}
\end{table*}


\subsection{Observation data and high-quality strong lens candidates}
\label{sec: real lens}
As anticipated, this paper focuses on two main datasets: KiDS and BELLS GALLERY. For this latter sample, we have specifically used the ones performed within the BELLS GALLERY sample as reference model performance. 
The main reason to choose KiDS is that this dataset provides some of the best image quality data (FWHM$\sim0.7''$ in $r-$band, see also below) currently available from ground observations and very similar pixel-scale to the one expected for the upcoming LSST observation. As such, we can use this dataset to prepare for the early LSST data strong lensing analyses. As we have discussed earlier, ground-based strong lensing systems are particularly challenging due to the higher mixing among the contribution of the light coming from the lens and the background light that introduces further degeneracies on top of the usual strong lensing model alone \citep{2024SSRv..220...12S}.
HST observations, on the other hand, due to their extraordinary image quality (typical FWHM$\sim0.1''$), provide a unique opportunity to fully validate our method, pushing it to the most extreme accuracy needed to optimize the image reconstruction of the lensing system and leave no residuals above the very low background of space observations.
Below, we summarize the data used and their parent observations.

\subsubsection{KiDS data}
The Kilo-Degree Survey (KiDS, \citealt{2013ExA....35...25D}) has been carried out using the OmegaCAM \citep{Kuijken2011OmegaCAM} mounted at the VLT Survey Telescope (VST; \citealt{Capaccioli2011Msngr.146....2C}).
It has covered an area of 1350 $\rm deg^2$ of the equatorial and southern sky in four ($ugri$) optical bands (see Wright et al. 2025 for a summary of the final data release observations). The field-of-view of OmegaCam is 1 $\rm deg^2$ with $0.2''/\rm pixel$ resolution (after final calibrated, coadded images), and the best image quality observations (FWHM$\sim0.7''$) obtained in $r$-band, used as reference filter for weak lensing observations. The optical observations from KiDS have been fully replicated within the VIKING project using the VIRCAM at the VISTA telescope \citep{Edge2013Msngr.154...32E} in the near-infrared (NIR) bands ($ZYJHK_S$).
To simulate lens images for ground-based telescopes such as KiDS, we needed to reproduce low SNR conditions and a noisy background realistically. 
To do that (see also \S\ref{sec: detail of mock}), the noise component is directly taken from KiDS survey images \Zhong{by randomly selecting noise ``cutouts'' centers in the KiDS footprint. In particular, we have selected 100\,000 of these cutouts by checking that there were no pathological situations\footnote{This has been done by checking the statistics of the mean and rms values of the pixel counts in the cutouts.} (e.g., stellar spikes, bright objects), but yet allowing for the presence of ``companion sources'' rather than using fully ``empty'' areas to consider the presence or random source in the field of view in training (see also \S\ref{sec: detail of mock} and last columns in Fig. \ref{fig: kids images}).}

As a test sample, on the other hand, to be used after the GGSL-UNet has been trained to check the performance for application to real data, we will use the high-quality candidates of real strong gravitational lenses from the catalogs of \citep{2017MNRAS.472.1129P, Petrillo2019a_MNRAS.482..807P, Petrillo2019b_MNRAS.484.3879P} \footnote{\url{https://www.astro.rug.nl/lensesinkids/full_left.html}} and \cite{Li2020ApJ...899...30L} derived from Fourth KiDS Data Release (DR4; \citealt{Kuijken2019A&A...625A...2K}). 
To select strong lens samples from KiDS, we follow two basic criteria. First, we keep the samples with the obvious structures of strong lensing cases, such as arcs or multiple images around the lens. Second, we select the cases matched with spectroscopic survey catalogs, such as SDSS \citep{Bolton2012AJ....144..144B, Smee2013AJ....146...32S, Dawson2016AJ....151...44D}, GAMA \citep{Hill2011_GAMA, Driver2011MNRAS.413..971D}, and MUSE \citep{2020ApJ...904L..31N}, which can provide precise spectroscopic redshifts (spec-z) for foreground galaxies. We finally identified 50 high-quality KiDS strong lens candidates, with their basic information shown in Table \ref{table:51_KiDS_lens}. This Table provides the $\it BPZ$ \citep{2000ApJ...536..571B,benitez2011} photometric redshift estimates based on 9-band ($ugriZYJHK_S$) photometric data from KiDS+VIKING, derived using the Gaussian Aperture and Photometry (GAaP; \citealt{gaap_Kuijken2008A&A...482.1053K}) flux method. It also lists the spectroscopic redshifts (spec-z) of the 50 candidates, obtained by matching to different spectroscopic surveys, in Table \ref{table:51_KiDS_lens}. The details of strong lensing analysis for the 50 cases will be presented in another paper (Luo et al. in prep) in the future. Here, we focus on the decoupling and photo-z estimation of foreground galaxies and arcs from background sources in this paper.

\subsubsection{HST data}
As space-based observations, we use the lenses from the Baryon Oscillation Spectroscopic Survey (BOSS) Emission-Line Lens Survey (BELLS) for the GALaxy-Ly$\alpha$ EmitteR sYstems (BELLS GALLERY, \citealt{Shu2016ApJ...833..264S}). This is chosen because it provides complete foreground and background modeling, which can be used to assess the performance of our GGSL-UNet. It consists of 17 $r$-band lens systems whose images will be used as test samples to compare with the GGSL-UNet reconstructed images.
The lenses in this sample have been identified using a spectroscopic method that detects the presence of anomalous emission lines from higher redshift galaxies superimposed on the spectra of luminous red galaxies (LRGs). The massive galaxies typically targeted by BOSS spectroscopy are at redshifts z$\sim$0.5-0.8. The emission lines, primarily [OII] $\lambda$3727, H$\beta$, and [OIII] $\lambda\lambda$4959,5007, originate from star-forming regions of background galaxies at higher redshifts (z$\sim$1-1.5) that are gravitationally lensed by the foreground LRGs. Regarding the modeling approach, we can benefit from the results of \citealt{Shu2016ApJ...833..264S} that have used the modeling tool $lfit_{gui}$ with a graphical user interface (GUI) to smoothly analyze the BELLS GALLERY lenses using HST F606W-band images with manual feature masks. The tool simultaneously fits the lens light (elliptical Sérsic profiles), the mass parameters (SIE with external shear), and source light (Sérsic or pixelated profiles) to the HST imaging data. A B-spline fitting is also applied to refine the background subtraction and source representation. Finally, an MCMC optimization has been applied to constrain the lens and source positions, flux ratios, and Einstein radii by minimizing the $\chi^2$  between the ray tracing model and the pixel data of the HST images. This approach balances efficiency with flexibility, yielding lens models and source reconstructions.

\subsection{Mock lens}
\label{sec: detail of mock}

As discussed in \S\ref{sec:real_or_mock}, we only use mock lens images as the training set. The flux of each pixel in a mock lens image is labeled as $I_i$, which is constructed from three different components:
\begin{align}
    I_i = \alpha I_f^\gamma +  \alpha \beta I_b + \alpha \beta \epsilon I_n,
    \label{eq:total_flux}
\end{align}
where $I_f$ represents the component of the foreground lens galaxy, $I_b$ represents the component of the background source, and $I_n$ represents the noise component sampled from sky survey images with empty centers. The Greek letters $\alpha, \beta, \gamma, \epsilon$ are the random values sampled from a uniform distribution during the training process, which is deemed as an online augmentation \citep[see also \S\ref{sec: Method} for more details]{shorten2019survey}.  
We stress here that Eq. \ref{eq:total_flux}, being defined pixel by pixel, encodes the fact that in lensing events, there is potential for a contribution from both the foreground and the background signal that cannot be sharply separated. This is prone to degeneracies among the different contributions that are harder to break for large spatial overlaps between the two components. This is the typical situation we have for ground-based observations with poorer seeing. 
We have tried to mitigate this situation in constructing our tool, as we will detail in \S\ref{sec: Method}.
The magnitude of the noise, $\alpha  \beta \epsilon$, ensures that the noise is statistically more minor than the foreground and background signals. The tuned three different components are defined as:
\begin{align}
    & F_f =\alpha I_f^\gamma, \\
    & F_b = \alpha \beta I_b, \\
    & F_n = \alpha \beta \epsilon I_n,
\end{align}
which are used as labeled images in training to identify the foreground, background, and noise pixels. It is necessary to ensure correct segmentation in the lens components.

The subscripts represent abbreviations for the image’s color band, row, and column. As an example, for a cutout image with four different wavelength bands $(u, g, r, i)$ and a size of $64 \times 64$, the image subscripts should be:
\begin{align}
    I_i = I_{\rm (band, row, coloum)} = I_{(4, 64, 64)}.
    \label{eq: dimensions}
\end{align}
Each band has corresponding randomly chosen magnitudes or slopes (represented by the Greek letters). Here, $\alpha = \alpha_{\rm (band, 1, 1)}$, and the other Greek letters are also 4-dimensional vectors sampled from a uniform distribution.
The range of the uniform distribution is shown in Table \ref{tabel: model paramters}. 
\Zhong{Note that the flat priors in these training parameters of Table correspond to a flat prior to the colors, which we have deliberately chosen because we wanted the network to reconstruct the lens based on its geometry rather than use the color information. This will ensure no color biases are introduced in the reconstruction of different bands.}

The foreground and background images in four different bands share the same ellipticity and effective radius. In real cases, the effective radius or Sérsic index might vary across different bands, which we approximate by tuning parameters such as $\alpha$ and $\gamma$.
The foreground models in different bands differ in magnitude and slope ($\alpha, \gamma$). Because the background source is a smaller, more compact object with a smaller effective radius, we did not consider its slope change for simplicity compared to the foreground. Therefore, the difference between the background models in different bands is only in magnitude ($\alpha, \beta$) and does not involve any slope ($\gamma$) tuning for simplicity.
We remind you that the noise is randomly sampled from the reference dataset (KiDS or HST in our case) with no source in the center but with no restriction on the presence of other sources off-center. This also allows us to include possible ``companion'' or ``contaminant'' sources in the training sample, reproducing realistic observational conditions that might or might not involve overlapping sources blended within the signal from the strong lensing components. 
The foreground and background images are normalized by their maximum values, while the noise is normalized by its median value.
Using these augmentations has two advantages. First, it can reduce the volume of training data. Second, the training data can explore a more expansive parameter space; the amplitude and slope parameters can tune the color information while the geometric information remains unchanged. 

To mock the foreground deflector and background sources, we have used a single Sérsic profile \citep{1999A&A...352..447C, 2003ApJ...582..689M}. Although this is not the most realistic model we can use to reproduce more complex multicomponent systems, this has been shown to capture most of the galaxy properties and provide a good approximation to produce training samples for deep learning tools that have successfully applied to real galaxies (see, e.g., the measurement of structural parameters, \citealt{2022ApJ...929..152L}). To generalize the typical strong lensing configurations for the foreground system, we have also considered a ``blending'' situation where twenty percent of the deflector is constituted by a pair of galaxies with two different Sérsic models and centers slight offset within one $\theta_E$. Furthermore, to train the U-Net to consider the cases of false detections, we randomly set 5\% of the training set made of $F_b=0$ (i.e., no-background signal) to help the U-Net to recognize cases without true galaxy background signal.

We used $\it{lenstronomy}$ \citep{2018PDU....22..189B} to simulate the foreground and the distorted background source. The (flux normalized) Sérsic model is defined as:
\begin{align}
    & I(\xi) = \exp \{ - b_n \xi^{1/n}\} \otimes PSF, \\
    & \xi = R/R_h, \ b_n = 2n - 1/3, \\
    & R = \sqrt{q x^2 + y^2/q}.
\end{align}

\begin{table}
    \caption{The parameters of the mocking lens. The top section shows the distribution of the augmentation parameters. The second part is the model of the foreground galaxy; the third part is the background source; the fourth part is the shear effect of the lens mass/potential model. The details of these parameters will be given in \S \ref{sec: detail of mock}. Considering the range of $\gamma$, the foreground effective $n$ should be approximately between 0.5 and 9 when the radius is around the half-light radius.}
    \centering
    \begin{tabular}{llll}
    \hline \hline
    Object & Para & Range & Unit \\ 
    \hline \hline 
    training & $\alpha, \beta, \epsilon$ & [0,1] & -- \\
             & $\gamma$   & [0.5, 1.5] & -- \\
    \hline \hline 
    foreground  & $x_l,y_l$ & [-0.6, 0.6] & arcsec \\
            & $R_{h}$ & [1.0, 3.0] & arcsec  \\
            & $n$ & [1, 6] & --   \\
            & $PA$ & [0, 180] & degree   \\
            & $q$ & [0.4, 1] & --  \\
    \hline \hline 
    background  & $x_{s},y_{s}$ & $ x_l, y_l \pm 0.8 \theta_E$ & arcsec   \\
            & $R_{h}$ & [0.05, 1.0] & arcsec  \\
            & $n$ & [1, 6] & -- \\
            & $PA$ & [0, 180] & degree  \\
            & $q$ & [0.2, 1] & -- \\
    \hline \hline
    lens    & $\theta_E$ & [1.0, 3.0] & arcsec  \\
            & $\gamma_1, \gamma_2$ & [0.0, 0.1] & -- \\
    \hline \hline
    \end{tabular}
    \label{tabel: model paramters}
\end{table}

\begin{figure}
        \centering
        \includegraphics[width=0.48\textwidth]{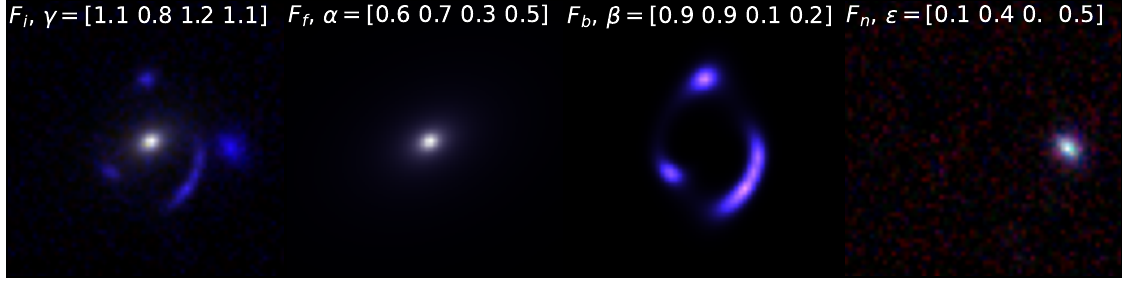}
        \includegraphics[width=0.48\textwidth]{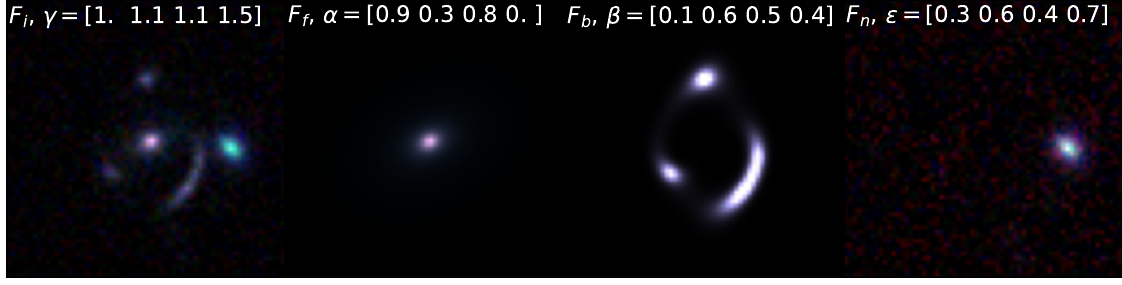}
        \includegraphics[width=0.48\textwidth]{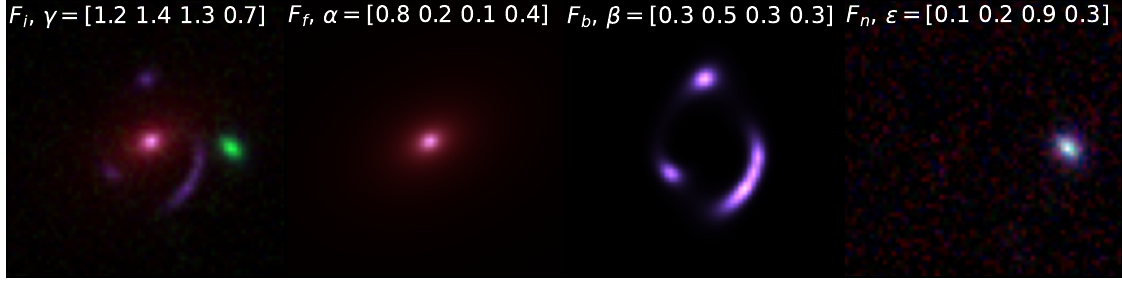}
        \includegraphics[width=0.48\textwidth]{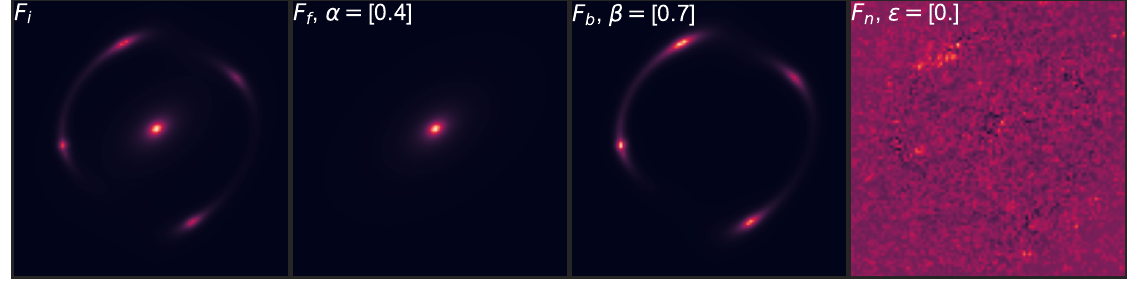}
        \includegraphics[width=0.48\textwidth]{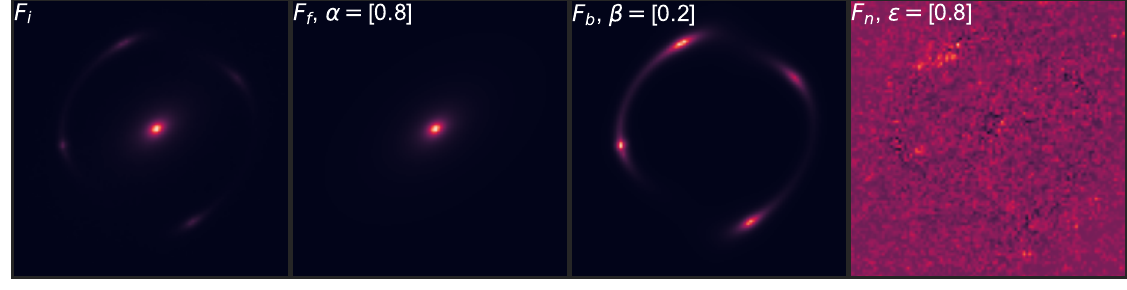}
    \caption{Examples of training/mock images. The first three rows are KiDS mock, and the last two rows are HST mock. Each KiDS mock image has a size of $4 \times 64 \times 64 $ pixels and a resolution of $0.2''/pixel$. The wavelength bands $(i, r, g)$ are mapped to the colors $(R, G, B)$. Each HST mock image is a size of $1 \times 101 \times 101 $ pixels and a resolution of $0.05''/pixel$.
    The 2nd column shows the mock foreground component, $F_f$; the 3rd column shows the mock distorted background, $F_b$; and the final column shows the noise cutout from the KiDS survey, $F_n$. By tuning the parameters $\alpha, \beta, \gamma, \epsilon$ during training (for the HST mock, we only tune its amplitudes because it is a single r-band image), the color changes, but the geometry remains the same. The 1st column, $F_i = F_f + F_b + F_n$, will be the input to the network, and the mock signal components $F_f, F_b$, as well as the survey cutout noise $F_n$ will be the outputs to be fitted.}     
    \label{fig: kids images} 
\end{figure}

\begin{figure}
    \centering
    \includegraphics[width=0.48\textwidth]{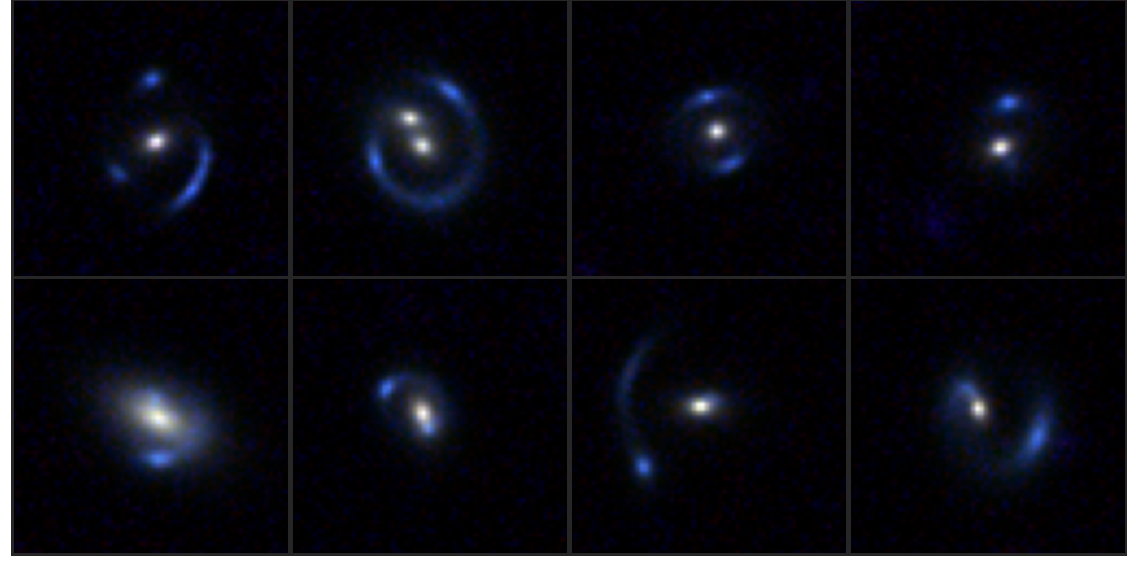}
    \includegraphics[width=0.48\textwidth]{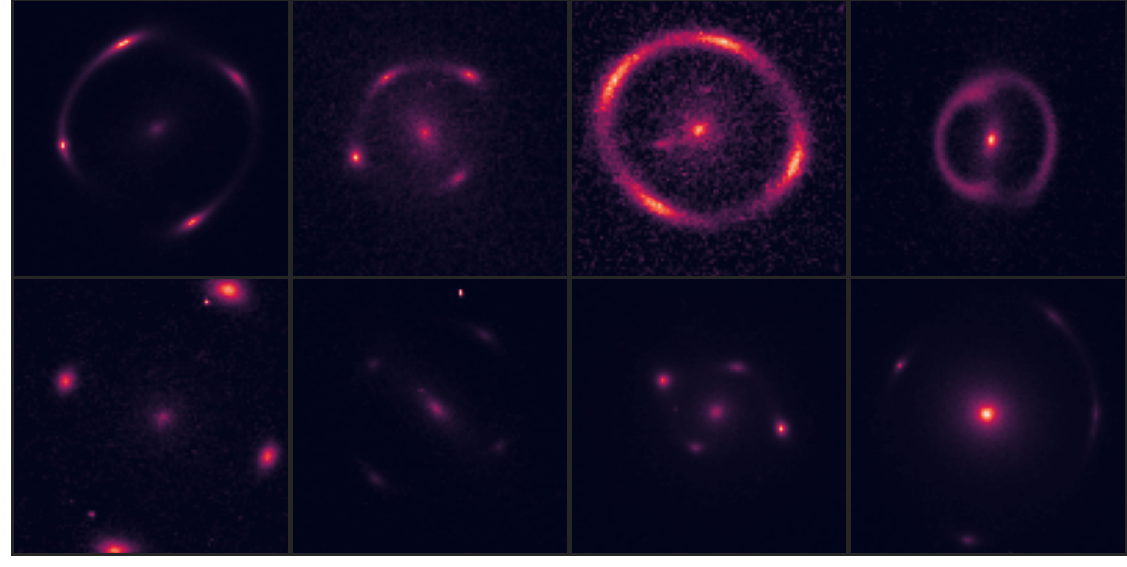}
    \caption{Examples of mock input images. The first two rows are the mock KiDS images $F_i$, while the last two rows are the mock HST images $F_i$.}
    \label{fig: mock_input}
\end{figure}

We can identify the following parameters: the radius $R$, from the center, $(x_l,y_l)$; the half-light radius, $R_{h}$; the position angle, $PA$; the Sérsic index, $n$; the axis ratio, $q$. The parameters for the foreground and the background S\'ersic profiles are shown in Table \ref{tabel: model paramters}. We used flat distributions for all parameters to avoid an unbalanced training sample and a realistic range of intrinsic parameters, as suggested from the real data (see, e.g., the BELL GALLERY, \citealt{Shu2016ApJ...833..264S}). 
As a first proof-of-concept test, in this paper, we cover a reasonable volume of the physical parameter space of the lensing systems, despite this might not fully consider all possible parameters, including extreme cases (for example, we are excluding very flat foreground systems with $q\sim0.2-0.3$). This option makes the network training more efficient, fast, and easier to converge while improving performance within the primary parameter space. \Zhong{Although the parameter space has been chosen to optimize the model training, the testing step in the following section shows that the GGSL-Unet still performs excellently in most mock/real lens test cases, as quantified by the chi-squares diagnostic.}
The projected potential of the background galaxy is calculated according to a singular isothermal ellipsoid profile (SIE, \citealt{1994A&A...284..285K}), which shares the same geometry parameters, like center, $PA$, and $q$, as the foreground Sérsic model. The convergence (or dimensionless surface density) for SIE is defined as: 
\begin{align}
    \kappa(x,y) = \frac{1}{2} \frac{\theta_E}{\sqrt{ q x^2 + y^2/q } },
\end{align}
where $\theta_E$ is the Einstein radius. The synthesis of the background galaxy will be sheared by large-scale external perturbations. This external shear is separated into two shear components, $\gamma_1$ and $\gamma_2$, along the radial and tangential directions. In Table \ref{tabel: model paramters}, we list the domain of these parameters in the bottom rows. 
The foreground and distorted background images are then convolved with a Gaussian Point Spread Function (PSF). For the KiDS-like lensing systems, we have adopted a constant $\sigma=0.4''$, corresponding to an FWHM $\sim 0.7''$, which is the average seeing for the $r$-band in KiDS and very close to the median seeing of the HQ strong lensing candidates in $g$- and $i$-band ($\sim0.8''$)\footnote{To save computational resources and storage, we used the same mock parameters/templates across different bands, applying the same $\sigma$ for all bands. For instance, we only mock 1x64x64 $I_f$ and $I_b$ with the PSF applied and then adjust the parameters $\alpha$, $\beta$, $\gamma$, and $\epsilon$ to produce images for the four bands. This approach reduces storage and memory usage by 75\%, which is beneficial when dealing with more bands (e.g., 9-band images) and large volumes of training data.}. Note that for this first experiment, our choice to ignore the (albeit small) band-to-band seeing variation can potentially impact the accuracy of the GGSL-UNet in real applications where a band-to-band seeing variation might be rather common. As we will test the trained network on real observations in \S\ref{sec: real lens}, we will keep this in mind as a possible source of systematics. However, we anticipate here that we will consider this implementation for future analyses (see \S\ref{sec: Conclusion}).
The PSF and the lower limit of $\theta_E$ are driven to optimize the performances on ground-based observations. 

Similarly, to predict BELL GALLERY lenses, we have also simulated HST $r$-band images, following the same procedure as for KiDS. The noise component is taken from real HST survey images as for KiDS, while the PSF is generated by Tiny Tim \citep{2011SPIE.8127E..0JK}, with a $\sigma$ of approximately 0.1$''$.

Some examples of mock KiDS-like and HST-like mock strong lensing configurations, generated under different augmentation parameters, are shown in Fig.~\ref{fig: kids images}.
They include the total lensing system images (first column on the left), the foreground galaxy images (second column), the background images (third column), and noise images (fourth column). The lensing system images are the input of our network, and the foreground, background, and noise images are the ones we want to reconstruct with the GGSL-UNet. The more configurations of mock input images $F_i$ are shown in Fig. \ref{fig: mock_input}.

Despite the simplified lens model used in our training data, we will show that the network can already reconstruct the lensing system's different components rather accurately, not only if applied to mock lens images but also to real strong lensing events.

\section{Method}
\label{sec: Method}

\begin{figure}
    \centering
    \includegraphics[width=0.5\textwidth]{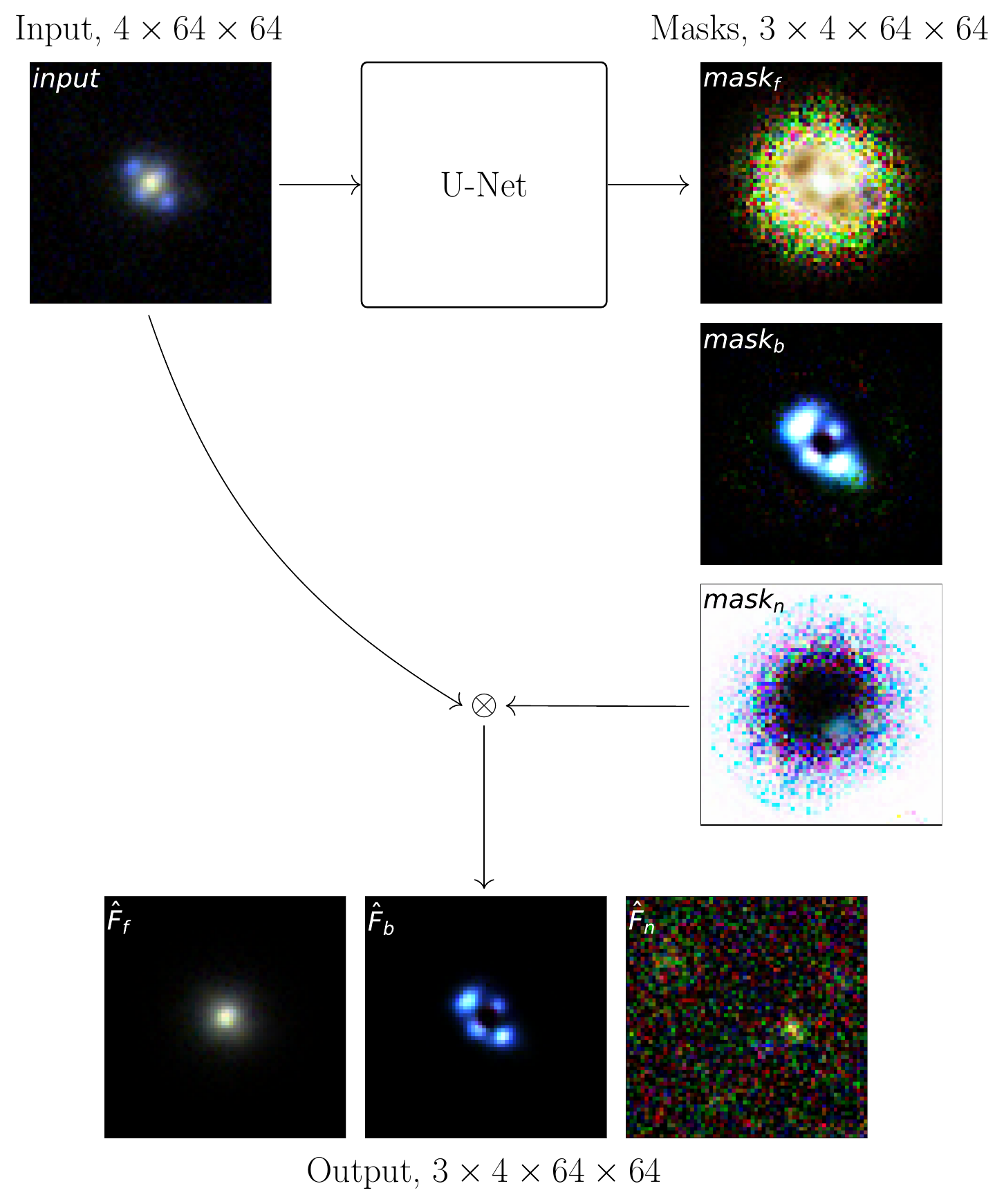}
    \caption{Masks generated by U-Net and the final output of different components, with a real KiDS lens image serving as the input. U-Net outputs three masks corresponding to three different components: the mask for the foreground, the mask for the background source, and the mask for noise. Masks can be considered as the probability of different components in each pixel. The notation $\otimes$ represents multiplication. The bottom three images show the final output of the three different components of the lens, which are the product of the input image and the respective mask.}
    \label{fig: U_Net_output}
\end{figure}

We use U-Net to extract the different components of lens images. 
The U-Net model is a type of generative model consisting of two main branches: down-sampling and up-sampling blocks with shortcuts at the same level. The down-sampling and up-sampling blocks mainly comprise ResNet structure units \citep{2015arXiv151203385H} and transformer structure units \citep{2017arXiv170603762V}.
\Zhong{The down-sampling reduces the dimension of features, while the up-sampling increases them. In our case, this means reducing the image's last two dimensions (see Eq. \ref{eq: dimensions}). For instance, the first down-sampling block in Table \ref{tabel: U-Net architecture} reduces the input image size from $4 \times 64 \times 64$ to $32 \times 32 \times 32$).}
We use the Python module \textit{diffusers} \footnote{\url{https://pypi.org/project/diffusers/}} \citep{von-platen-etal-2022-diffusers} to quickly build a U-Net model, in which the blocks and the channels used are shown in Table \ref{tabel: U-Net architecture}. 

U-Net has been widely applied in image segmentation \citep{2020arXiv201101118S}, producing output images of the same size as the input and effectively modeling noise through skip connections in its architecture. Recently, U-Net has also been applied to improve the performance of strong gravitational lens searching \citep{2024MNRAS.533.1426N} and modeling \citep{2022A&A...657L..14O, 2024arXiv240608442R}. 
Besides image segmentation, it can also be integrated into a framework to enhance the quality and accuracy of generated scientific images.

\begin{table}
    \caption{U-Net architecture. The first two columns are the downsample block and the corresponding number of channels, while the last two columns are the upsample block and the corresponding number of channels. 
    `DownBlock2D' or `UpBlock2D' consists of two 2-dimensional convolution layers with skip connections. `AttnDownBlock2D' or `AttnUpBlock2D'  consists of the transformer structure.}
    \centering
    \begin{tabular}{|l|l|l|l|}
    \hline \hline
    Down block & Channels & Up block & Channels \\ 
    \hline \hline 
    Input           & 4     &  Output/Masks  & 12 \\
    DownBlock2D     & 32    &  UpBlock2D     & 32 \\
    DownBlock2D     & 64    &  UpBlock2D     & 64 \\
    DownBlock2D     & 128   &  UpBlock2D     & 128  \\
    AttnDownBlock2D & 256   &  AttnUpBlock2D & 256 \\
    \hline \hline 
    \end{tabular} \label{tabel: U-Net architecture}
\end{table}

\begin{algorithm}[t]
\caption{Training Loop for U-Net.} \label{alg: training loop}
\KwIn{Training data $X_m=(I_f, I_b) \sim  U(a,b)$ and $X_r=I_n \sim Survey$, number of epochs $N$, batch size $B$, learning rate $\eta$}
\KwOut{Trained model parameters $\theta$}
\BlankLine
\For{$epoch = 1$ \KwTo $N$}{
    \ForEach{batch $(I_f, I_b, I_n)$ \textbf{in} $(X_m, X_r)$ \textbf{split into batches of size} $B$}{
        \BlankLine
        $\alpha, \beta, \gamma, \epsilon \sim U(a,b)$\;
        $F_f = \alpha I_f ^\gamma, \ F_b = \alpha \beta I_b, \ F_n = \alpha \beta \epsilon I_n$\;
        $F_i = F_f + F_b + F_n$\;
        $F_x = Norm(F_x), \ x=(i,f,b,n)$ \;
        \textbf{Forward pass:}\\
        $(\hat F_f, \hat F_b, \hat F_n) \leftarrow \text{U-Net}(F_i | \theta)$\;
        $loss \leftarrow L_1(\hat F_x, F_x)$\;
        \BlankLine
        \textbf{Backward pass:}\\
        $grads \leftarrow \text{ComputeGradients}(loss, \theta)$\;
        \BlankLine
        \textbf{Update parameters:}\\
        $\theta \leftarrow \theta - \eta \cdot grads$\;
    }
    \BlankLine
    \textbf{Evaluate on the validation set (optional):}
    $val\_loss \leftarrow L_1(\hat F_x, F_x)$\;
}
\Return $\theta$\;
\end{algorithm}
The input to the U-Net is preprocessed to have a uniform range of flux. The flux $F_x$ is normalized by the maximum flux within the input image's $1 \times 1 \rm arcsec$ center area as follows:
\begin{align}
    Norm(F_x) = F_x / {\rm max}(F_i(1'' \times 1'')).
\end{align}

To avoid fake outputs or ``hallucinations'' \citep{ji2023survey}, which is important for scientific applications, we will use some `masks' generated by U-Net. By multiplying the masks with the input images, we aim to produce realistic output images. Our adopted procedure is shown in Fig. \ref{fig: U_Net_output}. Here, the mask generated by the U-Net represents the probability of different components in each pixel. These probabilities are positive, by definition, and they sum up to 1 by normalization. With these definitions, the flux of each component of the total signal per pixel (i.e., foreground, background, and noise), as well as the normalization of the masks, is defined as:
\begin{align}
    & \hat F_x = F_i \times \rm mask'_x, \\
    & \rm mask'_x = \frac{mask_x^2}{mask_f^2+mask_b^2+mask_n^2}, 
\end{align}
where $F_i$ is the input flux of the image, $\rm mask_x$ denotes the mask, and $\hat F_x$ the predicted flux for the foreground (f), background (b), and noise (n), i.e., $x=f, b, n$, respectively. 

As a loss function to fit the output to the different components of the lens, we use the L1 function defined as:
\begin{align} \label{eq: total loss}
    L_1= \text{mean} (\sum_x \left| \hat F_x - F_x \right|),
\end{align}
where $ \hat F_x$ is the predicted flux value, and $F_x$ is the true value of different components. The sum of the mean absolute error is used because we expect to have the same back-propagation gradients for different components. 
We trained the network in the PyTorch \footnote{\url{https://pytorch.org}} framework.

\begin{figure}
    {\begin{minipage}{1.0\linewidth} 
           \includegraphics[width=1\textwidth]{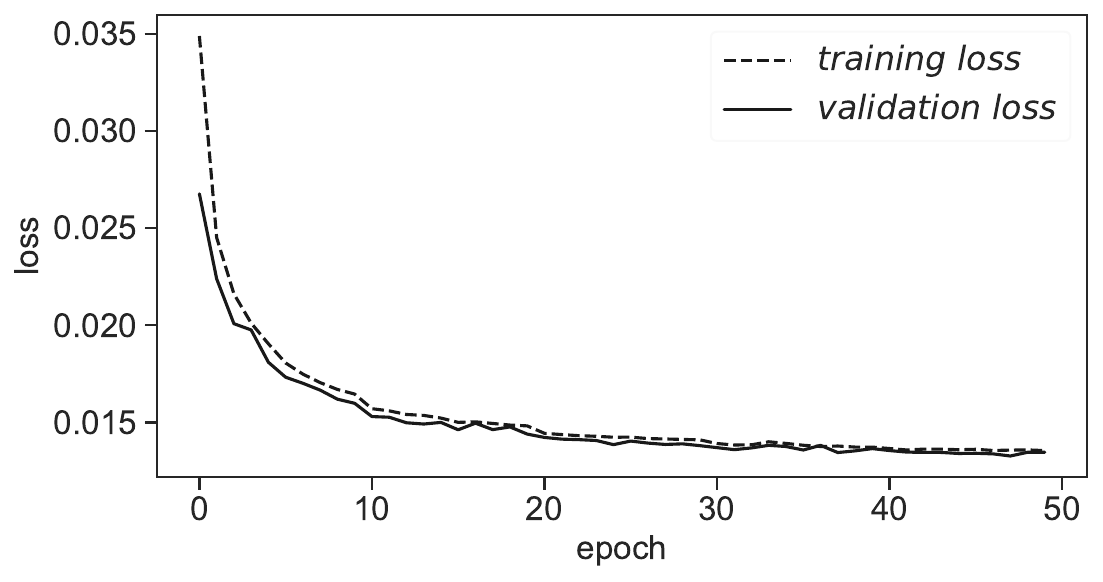}
    \end{minipage}}
    \caption{The loss of 50 training epochs. As the epochs increased, the loss of validation samples fluctuated around a constant value. The checkpoints with minimal validation loss will be used as our model. \Zhong{The test and validation data are both mock augmented data (90\% and 10\% of the total).}
    Due to the significant overfitting prevention by online augmentation, the training and validation losses appear close and overlap.} 
    \label{fig: total loss} 
\end{figure}

\begin{figure*}
    \centering
    \includegraphics[width=0.9\textwidth]{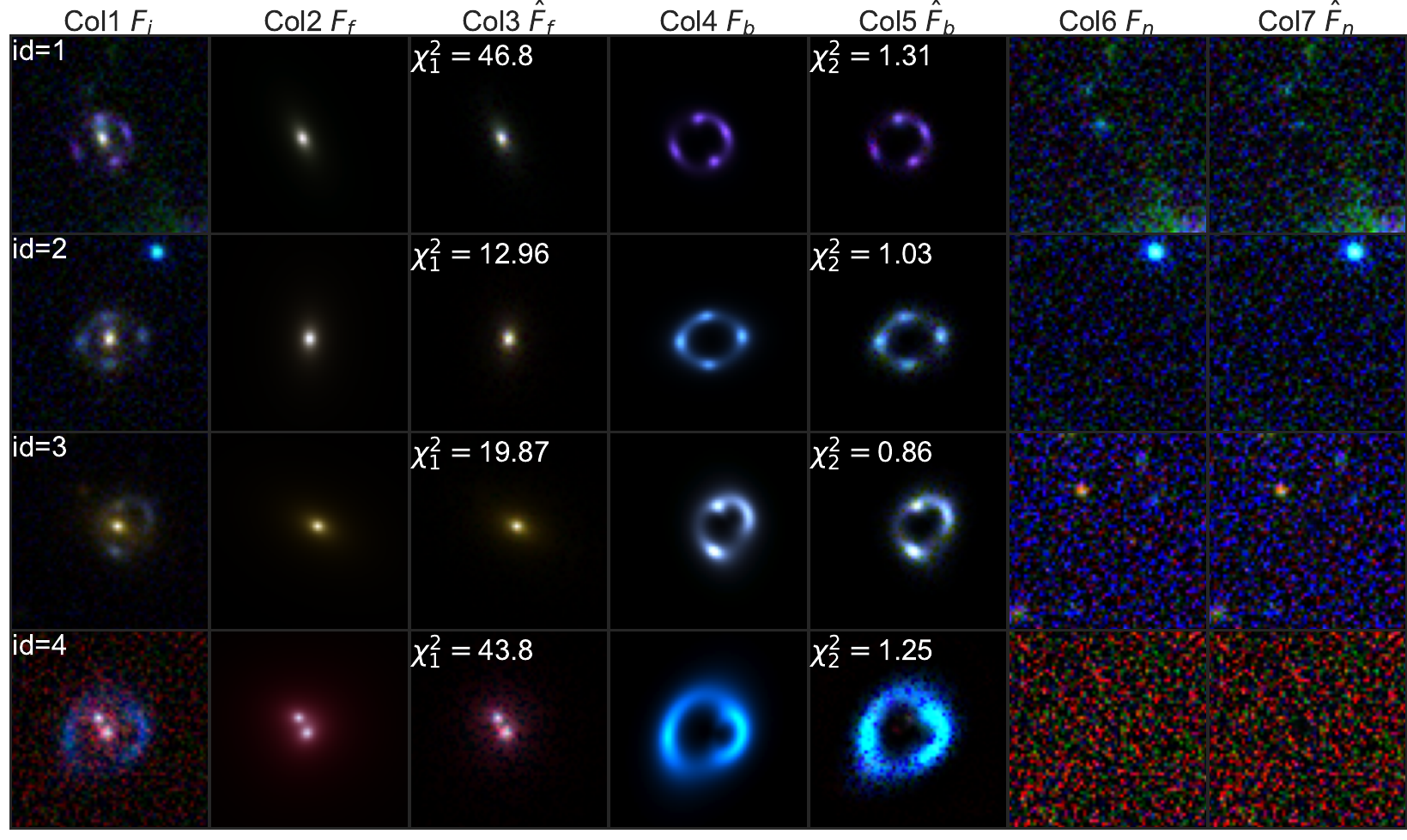}
    \caption{The mock KIDS images and their reconstructed images using U-Net are shown face-to-face. Each row represents one object with its different components and the corresponding reconstructed images. The input $F_i$ to the U-Net is a 4-band $64 \times 64$ pixel lens image, while the output of the U-Net includes the foreground galaxy $\hat F_f$, the background image $\hat F_b$, and the residual $\hat F_n = F_i - \hat F_f - \hat F_b$. 
    The mock foreground galaxy is $F_f$, and the mock background image is $F_b$. The last column shows the mock subtracted residual $F_n = F_i - F_f - F_b$.}
    \label{fig: kids reconstructed images} 
\end{figure*}

The training loop of the network is described in the Algorithm \ref{alg: training loop} inset. 
Here, the training sets consist of two kinds of data. First, the theoretical mock data, $X_m=(I_f, I_b) \sim  U(a,b)$, are obtained by sampling their parameters from the uniform distribution listed for the foreground, background, and lens components in Table \ref{tabel: model paramters}. 
Second, the real noise data, $X_r=I_n \sim Survey$, are randomly picked from observed images. 
Following the Algorithm \ref{alg: training loop} inset, for the mock data, we have sampled the parameters from the distribution $U(a,b)$ (last three parts of Table \ref{tabel: model paramters}), generated $I_f$ and $I_b$, and then sampled $I_n$ from the survey images. 
The model is trained in batches, where the augmentation parameters are sampled from $U(a,b)$ (the first part of Table \ref{tabel: model paramters}) to obtain $F_f, F_b, F_n$, and $F_i$ as the training targets. As discussed in \S \ref{sec: detail of mock}, we have used online augmentation. This means the augmented training samples are constructed ``on the fly'' during training. For instance, this can include image rotating or zooming. In this case, we change the magnitude, slope, and color during training.

A total of 100\,000 mock images are generated and separated into $90\%$ for training and $10\%$ for validation. Another new identical mock of 1,000 images will serve as a test set to demonstrate the model performed on mock data.
Fifty epochs were executed for the training. During this phase, the loss function in Eq. \ref{eq: total loss} is monitored to assess the model convergence.
The training and validation losses are plotted in Fig. \ref{fig: total loss}, where we can see that, after training, the loss converged to a stable value, and the checkpoints with minimal validation loss were saved.

\section{Evaluation and results}
\label{sec: Evaluation}
In this section, we discuss the performance of the GGSL-UNet on both mock and real data. Besides evaluating the accuracy in the segmentation of the different components of the lensing systems, we will discuss how to use the ability of the GGSL-UNet to efficiently pick any strong lensing feature and the background contribution (including nearby projected random galaxies) to classify the images as lensing or non-lensing systems. We also assess the accuracy in the total flux reconstruction of the different components to produce catalogs of multi-band total magnitudes of the foreground and background systems for spectral energy distribution analyses (e.g., photometric redshifts or stellar populations).

\subsection{GGSL-UNet as foreground/background/noise segmentation tool}
\label{sec:reconstruction}
Some examples of mock images and their reconstructed images by U-Net are shown in Fig. \ref{fig: kids reconstructed images} for the KiDS mock ``color'' images (we do not show the HST for brevity).
The images have three different components: the mock foreground galaxy $F_f$, the mock background source $F_b$, and the noise sample from the KiDS survey $F_n$. The corresponding parts reconstructed by U-Net are $\hat F_f, \hat F_b, \hat F_n$.
Here, the $\chi^2$ is used to evaluate the reconstruction quality, and the residuals allow us to visually assess the reconstructed results, as this should be close to the original ``noise'' cutout from the real data used to create the mock system.

To evaluate the fitting of different components from the U-Net segmentation and highlight the background signal contribution, we define two different (reduced) $\chi^2$ metrics:
\begin{align} \label{Eq: chi1}
    & \chi_1^2 = \sum_{p=1}^{N} \frac{(F_i - \hat F_f)^2}{\sigma_i^2}/N, \\ \label{Eq: chi2}
    & \chi_2^2 = \sum_{p=1}^{N} \frac{(F_i - \hat F_f - \hat F_b)^2}{\sigma_i^2}/N,
\end{align}
where $N = 4 \times 64 \times 64$, and $p$ represents each pixel of the image. 
For the mock images, we take $\sigma = |\hat F_n| + \epsilon$, where $\epsilon=10^{-3}$ to avoid divergence. Regarding the $\chi_2^2$, we remark that, by definition, the numerator corresponds to $\sim(F_n)^2$ if $\hat F_f \sim F_f$ and $\hat F_b \sim F_b$ (i.e., if the foreground and background fluxes are very accurately reconstructed as the true ones), while the denominator is $\sim ({F_n})^2$, hence $\chi_2^2\sim1$ if the reconstructed noise is almost identical to the true noise (provided that also the foreground and background images are well reconstructed, see above). In this respect, the $\chi_2^2\sim1$ is a measure of how well the GGSL-Unet can reconstruct the image noise, not only the overall strong lensing signal. Obviously, noise reconstruction is easier for ``empty'' fields with little contamination from other external sources (see Appendix \ref{app: failed reconstruction of mock}).

The $\chi^2$ values for each object are labeled in the corresponding panels in Fig. \ref{fig: kids reconstructed images}. These two different $\chi^2$ are calculated to highlight the contribution of the background source signal. For a lens image, one should expect $\chi_2^2$ to be smaller than $\chi_1^2$ as the former is calculated after the background is also fitted, rather than only the foreground. This is seen in Fig. \ref{fig: kids reconstructed images}.
\begin{figure}
    \centering
    \includegraphics[width=0.45\textwidth]{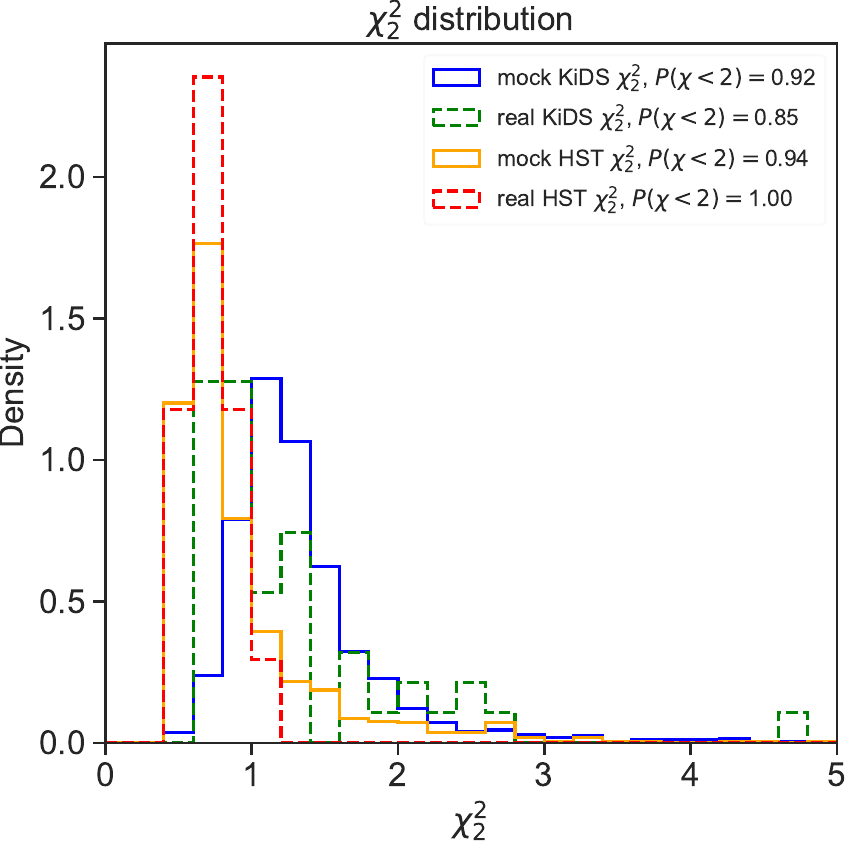}
    \caption{The $\chi^2$ density distribution of 1000 mock KiDS and HST strong lensing systems and the corresponding real data reconstructions. The $\chi^2$ values are binned with a width of 0.25. The proportion of cases where $\chi_2^2 < 2$, denoted as $P(\chi<2)$, is shown on the label.}
    \label{fig: kids_chi2_hist}
\end{figure}
Indeed, in the figure, we can see that the U-Net can recover the geometry and color information of the foreground and background components, which usually produces $\chi^2_2<2$. In these cases, by visually comparing the residual images, $F_n$ and $\hat F_n$, we can see that the U-Net can reproduce the same level of noise as in the original data from the corresponding survey, demonstrating that we can extract information by properly modeling the real noise. Strikingly, the network can reconstruct very complex cases as pair-system lenses with a similar level of accuracy (see bottom row of Fig. \ref{fig: kids reconstructed images}, but see also Appendix \ref{app: failed reconstruction of mock}).
On the other hand, we always have $\chi^2_1 >\chi^2_2$ due to the presence of the background signal in the $F_i - \hat F_f$ in Eq. \ref{Eq: chi1}. This offers an opportunity to separate strong lensing from non-strong lensing events: indeed, by comparing $\chi^2_1$ and $\chi^2_2$, one can expect that reconstructions with $\chi^2_1 \sim\chi^2_2$ would correspond with a negligible contribution from background signal (either from faint source or no-source at all). We will discuss the GGSL-UNet as a tool for lens finding in \S\ref{sec:lens_find}. 

Going back to the performances on the mock samples, to quantify the overall level of accuracy reached by the GGSL-UNet on the mock (KiDS and HST) samples, we show the distributions of the reduced $\chi^2_2$ returned by the GGSL-UNet for the KiDS-like and HST-like mock test samples in Fig. \ref{fig: kids_chi2_hist}. Here, we can see that the $\chi^2_2$ distribution of both KiDS-like and HST-like mock lenses peak at $\chi^2_2\sim1$, with the performance of the GGSL-Unet on HST systematically better than on KiDS. The overall $\chi^2_2$ distribution is also sharper for HST, with 90\% of the models showing a $\chi_2^2 < 2$ against over 80\% of the KiDS images. This demonstrates that we can successfully reconstruct the majority of mock images from ground and space. However, the latter allows the U-Net to obtain a better accuracy due to the better image quality (see, again, the discussion at the beginning of \S\ref{sec: detail of mock}).
For the systems with $\chi^2$ ($>2$), the GGSL-UNet ``poorly performed'' or even ``failed'' to reconstruct the strong lensing systems with reasonable accuracy (some large $\chi_2^2$ are attributed to the `companion sources’) even for the space images. These cases are discussed in Appendix \ref{app: failed reconstruction of mock}.

\begin{figure}
    \includegraphics[width=0.48\textwidth]{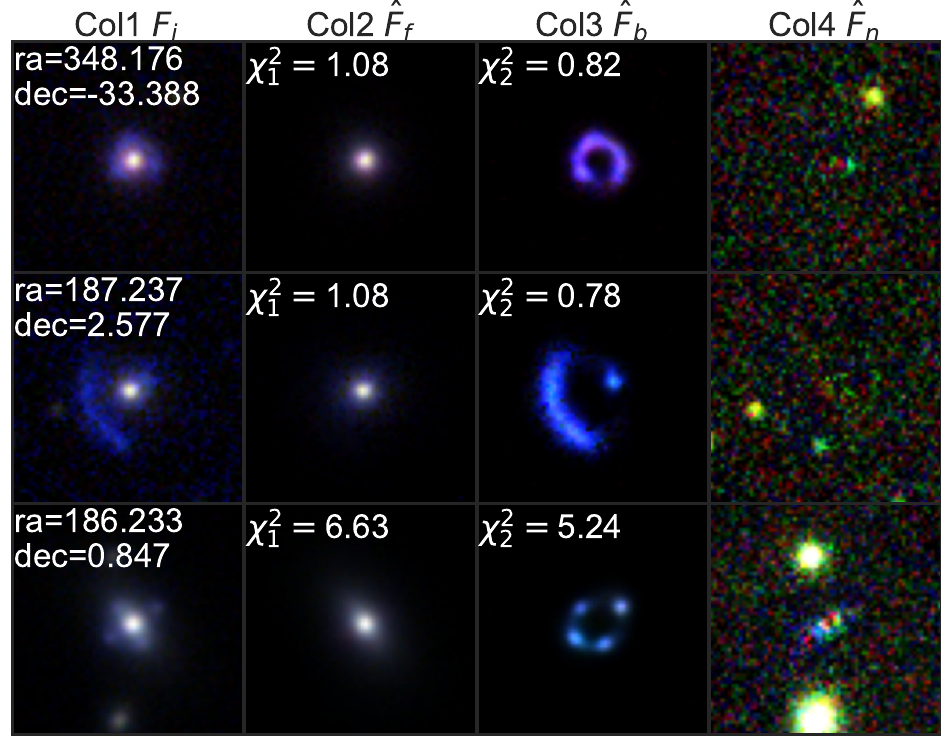}
    \includegraphics[width=0.48\textwidth]{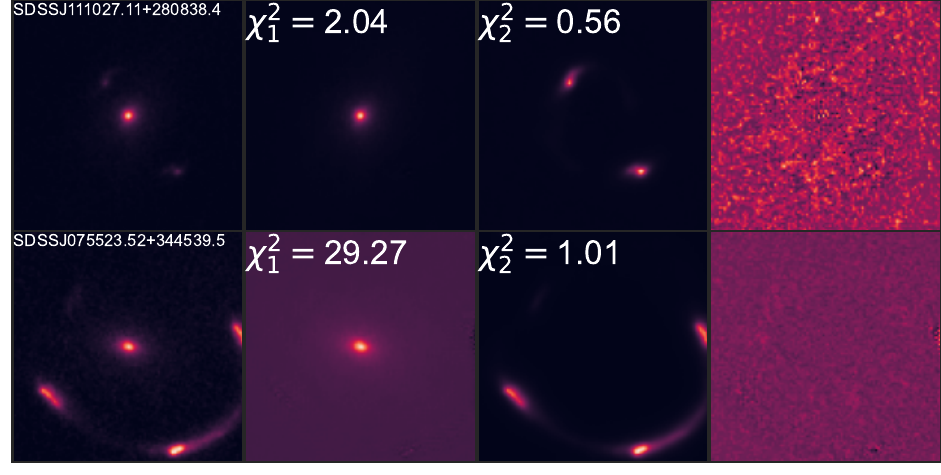}
    \caption{The real KiDS/HST images and their reconstructed images using U-Net. Each row represents one object with its different components' reconstructed images. 
    The input $F_i$ to the U-Net is a 4-band $64 \times 64$ pixel KiDS lens candidates image or r-band $101 \times 101$ pixel HST lens images, while the output of the U-Net includes the foreground galaxy $\hat F_f$, the background image $\hat F_b$, and the residual $\hat F_n = F_i - \hat F_f - \hat F_b$.}  
    \label{fig: real kids reconstructed} 
\end{figure}

We have finally tested the GGSL-UNet on the real lensing systems. In Fig. \ref{fig: real kids reconstructed}, we show some examples of the reconstruction of the foreground/background systems and the related residuals (more examples are shown in Appendix \ref{app: More real KiDS lens}). 
In the figure, we can see the U-Net reconstruction of the two lens components, $\hat F_f$ and $\hat F_b$, as well as the residual $\hat F_n$ images that, again, generally show a typical noise structure, including the companion/contaminant objects in the field-of-view left untouched. This illustrates how well the mock lens and the real lens foreground and background systems can be reconstructed. We stress that by properly modeling noise with U-Net, we can accurately extract the physical information of the deflector and source that are useful for science purposes (see, e.g., \S\ref{sec:photometry_reconstr}).

As for the mocks, also for the real systems, we can use the $\chi^2_1$ and $\chi^2_2$ metrics as in Eqs. \ref{Eq: chi1} and \ref{Eq: chi2}, where $\sigma$ represents the true photometric errors ($\sigma^{-2}$ is the inverse variance).
The $\chi^2$ values for each object are labeled in the corresponding panels in Fig. \ref{fig: real kids reconstructed} for both KiDS and HST. In the real case, we also find generally $\chi^2_1 > \chi^2_2$, a difference that looks larger for larger constraint strong lensing events (as anticipated earlier), as the $\chi^2_1$ is dominated by the bright residuals of bright source images left behind after the foreground galaxy has been subtracted.

To summarize the accuracy of the total sample of reconstructed ``real strong lensing events,'' in Fig. \ref{fig: kids_chi2_hist}, we also show the $\chi_2^2$ distributions of the reconstructed real lenses. Interestingly, both the real KiDS lenses and the HST lenses follow the $\chi_2^2$ distributions of the corresponding mock samples, meaning that the quality of the reconstruction of the real lenses is similar to the one of the mock lenses. This remarkable result suggests that the GGSL-UNet is applicable to real data with no degradation of the reconstruction power despite being trained on ``simulated'' strong lensing systems (see \S\ref{sec: Method}).

\subsection{GGSL-UNet as lens finding tool}
\label{sec:lens_find}
We have anticipated in the previous section that comparing the $\chi_1^2$ and $\chi_2^2$ values after the reconstruction provides a possible way to detect new lens objects. The way the GGSL-UNet would work here is that it would be fed with cutouts of a generic data sample, obtaining the majority of normal ``unlensed'' systems, with just a foreground galaxy for which the GGSL-UNet would produce a $\chi_1^2$, $\chi_2^2$ values from the reconstruction of the foreground and the background component. 

If we define the difference between $\chi_1^2$, $\chi_2^2$ as $\bar \chi^2$, we can classify the lensed and unlensed cases as follows:
\begin{align} \label{eq: chi1_chi2}
    & \begin{cases} 
        \chi_2^2 \sim 1 \\
        \bar \chi^2 = \chi_1^2 - \chi_2^2, \\
        \bar \chi^2 \approx 0 & \text{unlensed}, \\
        \bar \chi^2 \le thr & \text{unlensed}, \\
        \bar \chi^2 > thr & \text{lensed},
     \end{cases}
\end{align}
where $thr$ is a threshold that depends on the SNR of the image.

\begin{figure}
    \centering
    \includegraphics[width=0.48\textwidth]{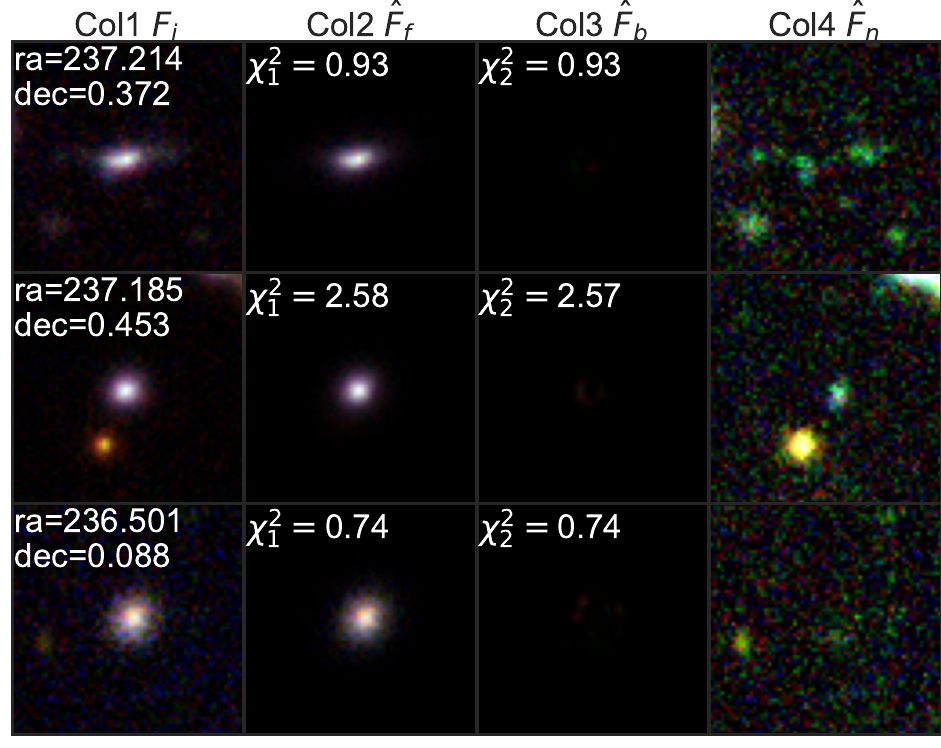}
    \caption{The real KiDS images and their reconstructed images using U-Net for unlensed cases are shown. These targets are randomly selected normal galaxies from the KiDS catalog. There are no signals in $\hat F_b$. Labels are identical to Fig. \ref{fig: real kids reconstructed}.}  
    \label{fig: real no lens kids reconstructed} 
\end{figure}

\begin{figure*} 
        \centering
        \includegraphics[width=0.45\textwidth]{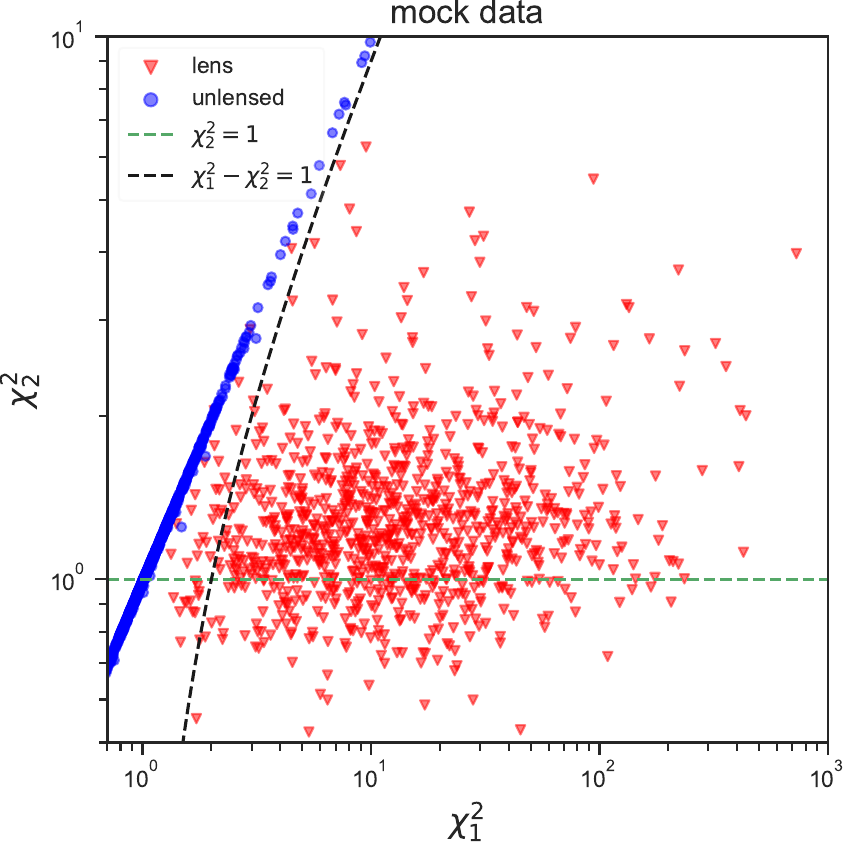}
        \hspace{10mm}
        \includegraphics[width=0.45\textwidth]{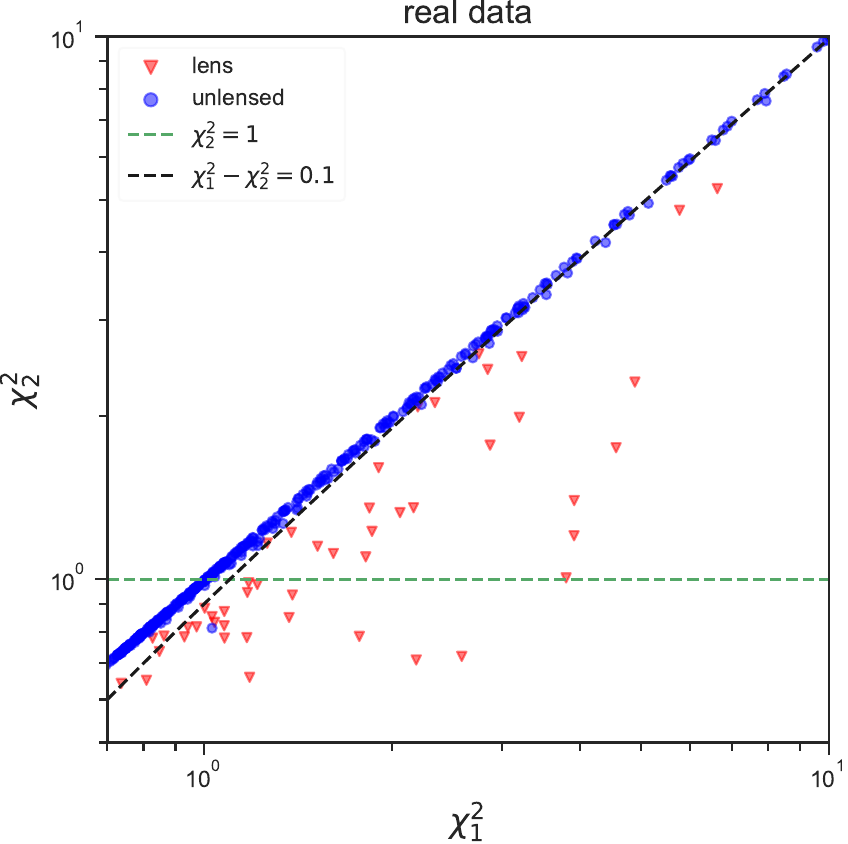}
    \caption{$(\chi_1^2, \chi_2^2)$ plane of mock data (left) and real data (right). The lines $\chi_2^2 = 1$ and the criteria $\chi_1^2 - \chi_2^2 = thr$ are plotted, as shown in Eq \ref{eq: chi1_chi2}. Due to the weaker background signal in the real sources, the $\chi_1^2$ values in the real data are smaller than those in the mock data. Therefore, we take $thr = 1$ for the mock data and $thr = 0.1$ for the real data as a demonstration.}    
    \label{fig: chi plane of mock} 
\end{figure*}

To demonstrate how the GGSL-UNet quantitatively would react to unlensed cases, we have also applied it to a sample of normal objects. Since we only generated mock lensing systems with an elliptical foreground galaxy, we have selected some real galaxies that looked like normal ellipticals. As shown in Fig. \ref{fig: real no lens kids reconstructed}, for these cases, there is no signal in $\hat F_b$, and $\bar \chi^2 \approx 0$, as expected.

\begin{figure}
    \centering
    \includegraphics[width=0.44\textwidth]{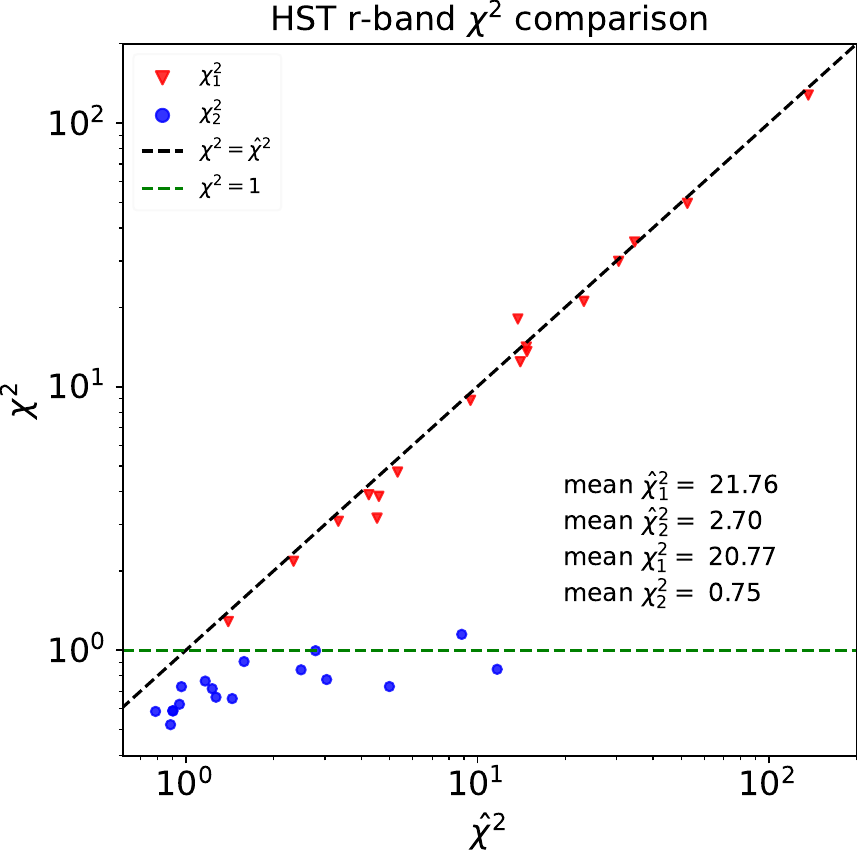}
    \caption{The distribution of $\chi^2$ for the 17 HST $r$-band images is presented. The $\hat{\chi}^2$ represents the chi-square calculated using the  BELLS GALLERY model, while the $\chi^2$ represents the chi-square calculated using the U-Net output.}  
    \label{fig: HST chi2} 
\end{figure}

\begin{figure*}
    \includegraphics[width=0.9\textwidth]{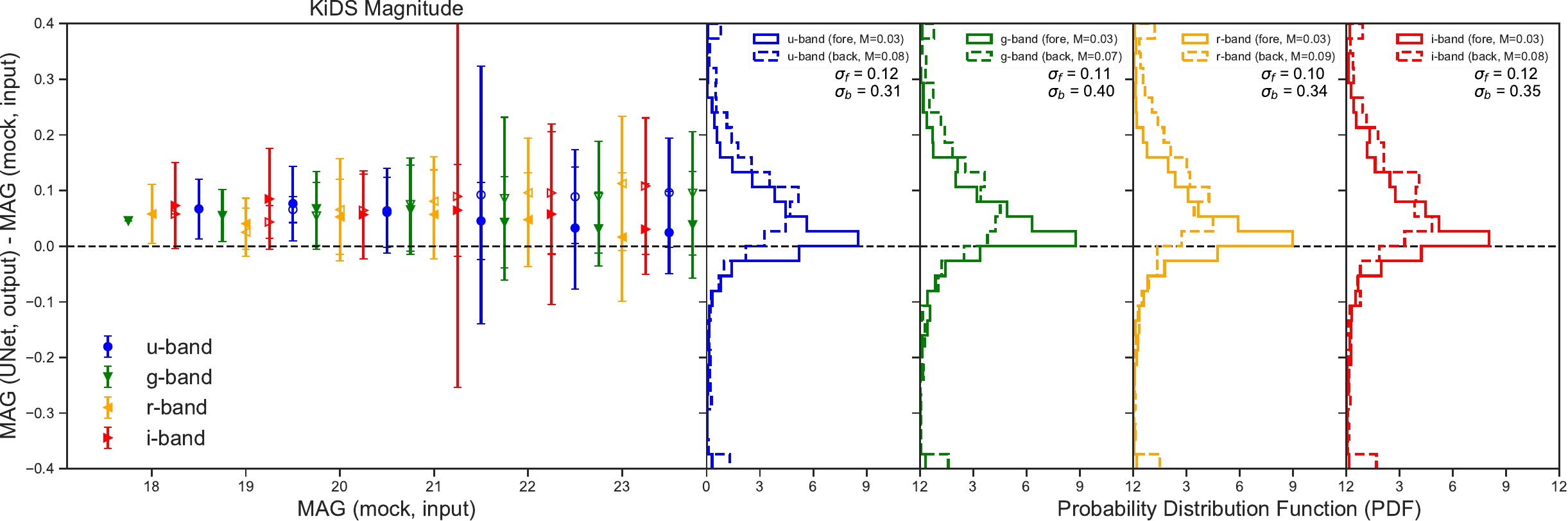}\\
    \hspace{-6.0cm}
    \vspace{0.5cm}
    \includegraphics[width=0.532\textwidth]{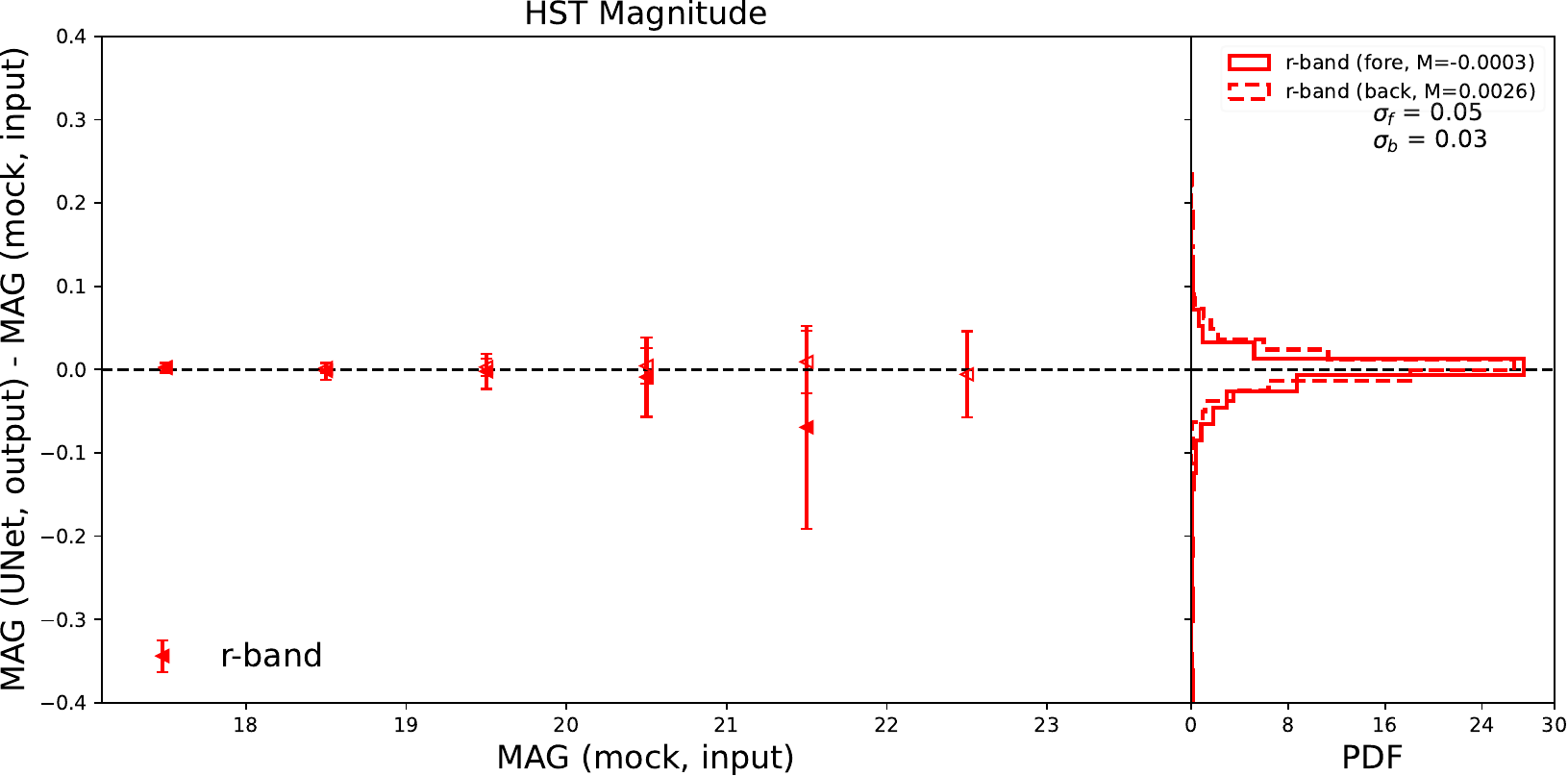}
    \hspace{10pt}
    \caption{The mean error of the foreground/background 4-band magnitudes for 1,000 mock KiDS and HST images in each bin, with a bin size of 1, is shown in the left panel. The foreground magnitudes are labeled with filled labels, while the background magnitudes are labeled with empty labels. The error bars represent one standard deviation. The density distributions of the predicted magnitude residuals with M (Median) values for the mock images are also presented on the right. In the left panel, the x-axis represents the magnitudes calculated from $F_f$ or $F_b$, while the y-axis represents the difference between the magnitudes calculated from $\hat F_f$ or $\hat F_b$ and those calculated from $F_f$ or $F_b$.}  
    \label{fig: mag of mock} 
\end{figure*}

To illustrate further the difference between lensed vs. unlensed systems, in Fig. \ref{fig: chi plane of mock}, We plot the $(\chi_1^2, \ \chi_2^2)$ plane for both mock data and real KiDS data.
In the left panel, we show the comparison of the $\chi_1^2$ and $\chi_2^2$ values for 1,000 mock lensed and unlensed. In the right panel, we take the 50 KiDS lens candidates and 500 randomly selected sources from the KiDS survey.
As expected from Eq \ref{eq: chi1_chi2}, $\chi_2^2$ should be around 1, and $\chi_1^2 \approx \chi_2^2$ for unlensed images, as shown by the line-dashed envelop in the figure demonstrating a very small scatter (obviously, $\chi_1^2$ is much larger than  $\chi_2^2$ for the lensing systems).
Due to the higher SNR in the mock data compared to the real data and the weaker background signal in the lens candidates, the $\chi_1^2$ values in the real data are much smaller than those in the mock data, but still significantly deviating from the 1-to-1 relation.

For completeness, in Fig. \ref{fig: HST chi2}, we have compared the results of the U-Net vs the the BELLS GALLERY ray-tracing model for the HST real galaxies face-to-face. This offers the opportunity to check the GGSL-UNet performances against some standard tool (as in \citealt{Shu2016ApJ...833..264S}). Note that, here, we are not asking the GGSL-UNet to derive the lens model parameters (the Einstein radius, axis ratio mass density, etc.) as the ray-tracing models do, but just compare the ``residuals'' after the image reconstruction from the U-Net using the same $\chi^2$ metrics for both the U-Net and the BELLS GALLERY ray-tracing model. 

The figure shows that the $\chi_1^2$ of the two methods is similar, i.e., that they model the foreground galaxy with similar accuracy. However, on the left-behind signal (i.e., the background + noise after the foreground subtraction), the GGSL-UNet largely outperforms the standard ray-tracing modeling, keeping for all systems a $\chi_2^2\sim1$ or smaller. The BELLS GALLERY sample has a mean $\chi_2^2 = 2.70$ (after the full model subtraction), while the GGSL-UNet has a mean $\chi_2^2 =  0.75$.
The reconstructed images and residuals are compared in Appendix \ref{sec: reconstruction of HST} for some systems for a visual comparison. This result is likely due to the ability of the U-Net to separate the background signal from the noise without being dominated by the model errors and degeneracies.
This opens two very promising perspectives: 1) that the generative models, or similar deep learning networks, can be used to more efficiently isolate the background signal for more accurate modeling using standard techniques, or 2) that one can use the GGSL-UNet itself to infer the main lens parameters. 

\subsection{GGSL-UNet as photometry reconstruction tool}
\label{sec:photometry_reconstr}
In image segmentation or de-noising, an exquisite $\chi^2$ does not guarantee that the signal segmentation (as in Eq. \ref{eq:total_flux}) is degeneracy-free. Since one of the great advantages of our method is the possibility to integrate the signal associated with every component to derive total fluxes, we need to demonstrate that the U-Net output images can preserve the flux information in every band and, consequently, the color information in multi-band catalogs. This is of paramount importance for science applications. One classic example is the derivation of the photometric redshifts, where the multi-band photometry and colors are fitted to theoretical or observational templates of galaxies to derive the redshift (see, e.g., $\it LePHARE$, \citealt{1999MNRAS.310..540A, 2006A&A...457..841I}). We will show such an application to the KiDS photometry derived with GGSL-UNet in Appendix \ref{sec: Downstream task}.
Another example is using the multi-band photometric information of the source and the deflector to derive stellar population parameters (see, e.g., \citealt{Li_6nuggets_arXiv}). Hence, we need to investigate the photometry accuracy that one can derive from the flux breakdown in all image pixels as described in \S\ref{sec: Method}. Obviously, this can only be done for the mock samples for which the `ground truth' magnitudes are known. We remark here that, as we use real noise background cutouts, we expect these, combined with the lensed source signal, to dominate the photometric errors, thus providing a realistic assessment of the overall precision of the reconstructed magnitudes. Also, we concentrate only on the KiDS sample for two reasons: 1) it is the only dataset for which we have the color information  (the BELL GALERY sample has only one band), and 2) it is the worst-case scenario as we have seen that, for HST, the reconstruction is always better so we can use the ground-based observation as an upper limit on the photometric accuracy reachable for the strong lensing reconstruction with GGSL-UNet. The magnitudes of the input and output mock images are then simply defined as\footnote{\url{https://kids.strw.leidenuniv.nl/DR4/format.php}}:
\begin{align}\label{eq: magnitude}
    m_x = -2.5 \log_{10} \sum_{\rm (x,y)=(0,0)}^{(64,64)} \hat F_x,
\end{align}
for every component $x= f, b, n$ as discussed in \S\ref{sec: Method}. Fig. \ref{fig: mag of mock} top panel shows the accuracy for the four band magnitudes of the foreground and background components of the KiDS mocks, using this definition, while the bottom panel shows the same results for the HST mocks. \Zhong{The x-axis represents the magnitudes for the mock (precisely $F_f$ and $F_b$ for the full or empty symbols, respectively), while the y-axis represents the difference between the output and input magnitudes for $\hat F_f$ or $\hat F_b$ (symbols as above). In the same figure, we show the distributions of the input-output magnitudes that allow us to highlight the presence of systematics. From this figure, we can derive two main features. 

First, looking at the input-output distributions, for KiDS, there are almost no systematics for the foreground magnitudes (median $M_f\sim0.03$ and scatter $\sigma_f\sim0.11$ across the different bands, as in the inset), although we also observe a tail toward the positive deviation (overestimate). Such an overestimate is more evident for the background magnitudes: here, the offset has a median $M_b\sim0.08$ and scatter $\sigma_b\sim0.35$.
For HST, there is no bias for both foreground and background magnitudes ($M<0.003$, i.e., smaller than typical photometric calibration errors) with a scatter that is $\sim80\%$ smaller than the corresponding KiDS $r$-band for the background magnitude, while it is only half of the precision of KiDS foreground photometry. 

Second, looking at the input-output vs. input magnitude plot, for KiDS, the foreground galaxies have magnitudes that are recovered within 1$\sigma$, especially for mag$>21$, and the median deviation is always contained within $\sim$0.05 mag in all bands (except for very bright sources at mag$<19$ that show a positive offset marginally consistent with zero within 1$\sigma$). The background source magnitudes, instead, are systematically overestimated by $\sim$0.1 mag. At a very close look, one may also see a trend to increase the offset for the background magnitude and decrease the offset for the foreground magnitudes toward fainter magnitudes (see comment below). 
There are no trends for HST, and all magnitudes are very accurately recovered by the U-Net. This shows a superiority of the predictions in space observations vs. ground-based observation in general, which might come from the better image quality (see below). 

In principle, one can foresee that the presence of these systematics for the background photometry can impact the quantities that are based on total fluxes (luminosities, stellar masses, etc.) as the fainter systems (typically higher redshifts or less magnified galaxies) would result fainter and less massive than they really are. As this trend with the magnitudes is similar in all bands, this does not seem to affect the colors. Hence, one can argue that these systematics would have less impact on the quantities based on colors (e.g., the photometric redshifts). 

Besides registering these potential systematics, we should try to assess the origin of these offsets here to elaborate strategies for future improvements. 
In this respect, the fact that both foreground and background systems are overestimated, regardless of the intrinsic magnitude, means that there is some flux missing from the U-Net segmentation as in Eq. \ref{eq:total_flux}. While one can expect that this should be easy to predict in the high-SNR pixels, this might become difficult at the low-SNR, where the signal is overwhelmed by the noise. In these pixels, we can argue that the GGSL-Unet tends to overweight the noise and subtracts flux to either the background, foreground, or both. As a partial demonstration of that, we have tested if the offset is alleviated using sharper surface brightness profiles for both foreground and background, setting the $n$-index to stay within [1-4] and limiting the $q>0.4$ to reduce the degeneracies among the parameters in the reconstruction. In this case, the offset for the foreground and the background systems reduced to $0.01$ and $0.07$ with a scatter of 0.1 and 0.2, respectively, hence suggesting a noticeable improvement. Currently, to keep the parameter space as wide as possible, we decided to live with these systematics for the KiDS reconstructed photometry, keeping in mind that they have to be related to the structure of the noise in KiDS.}
Indeed, this effect is not seen in the HST images (see bottom panel of Fig. \ref{fig: mag of mock}). According to the discussion above, we can explain this with the fact that, first, for HST images, the SNR is much higher than in KiDS and, second, the PSF in HST is sharper, and then the quality is higher than in the KiDS observations, making even the smoother light profiles easier for the U-Net to detect and then reconstruct the related image.

\section{Discussion and conclusion}
\label{sec: Conclusion}

In this paper, we have applied the generative network U-Net to reconstruct different components of galaxy-galaxy strong lensing images, namely the foreground ($f$) deflector light, the deformed image of the background source ($b$), either arc-like or multiple-imaged, and the noise ($n$), usually denoted as ``sky background.'' The tool for this Galaxy-Galaxy Strong lensing reconstruction using U-Net (GGSL-UNet for short) has been applied to high-quality strong lensing candidates from the Kilo Degree Survey (KiDS) and for a sample of HST-confirmed strong lensing events fully modeled within the BELLS GALLERY project.

The GGSL-UNet's main feature is the generation of masks that represent different ``weights'' for the contribution from the foreground, background, and noise to each pixel’s flux.
The network has been trained on realistic training samples that we have constructed by combining noise (i.e., sky background) samples from different real survey data with theoretical mock foreground and background models, assumed to follow S\'ersic profiles, which we have further improved by taking advantage of online augmentation.
Using an appropriate $L_1$ loss function to guide the U-Net in fitting different components  —- especially the noise -— in the training data, we have optimized the performances of the U-Net in extracting the signals for the foreground and background systems. 
Despite the fact that the light densities from the $f$/$b$ components are rather idealized (single S\'ersic profile, no substructures, etc.), the U-Net could still learn how to accurately model the real noise, allowing the network to extract these signals correctly. 

By modeling the multi-color band images of KiDS, we have demonstrated that a well-trained U-Net network can reproduce the foreground and background components for both mock and real data. The color and geometry information in the images are well retained in the reconstructed components. The U-Net performs well even in low SNR situations, with a $\chi^2$ (defined in \S\ref{sec:lens_find}) remaining within reasonable values. For cases with no strong lensing signatures, the U-Net could still model the foreground source very effectively, indicating that the generative network can also be used as a source extraction tool (see, e.g., {\it SExtractor}, \citealt{bertin1996sextractor}).

The GGSL-UNet used to reconstruct different components of the HST $r$-band images differs from the one used for KiDS strong lenses, mainly for the noise component and the different PSF adopted to convolve the S\'ersic models of both the foreground (deflector) and the background (source) galaxies.
We have compared the reconstructed components with the BELLS GALLERY models (see Appendix \ref{sec: reconstruction of HST}) and found that the U-Net reconstruction shows an accuracy comparable to classical methods. 
Comparing the mean $\chi_2^2$, we demonstrate that GGSL-UNet has a better overall performance in reconstructing the total luminosity from the two strong lensing components, foreground+background, as shown in Fig. \ref{fig: HST chi2}.

In addition to the segmentation of strong lensing events, we have also shown other possible applications of the network to other typical tasks to be performed in large imaging surveys like denoising images and extracting sources as illustrated in Fig. \ref{fig: real no lens kids reconstructed} and searching for lenses as shown in Fig. \ref{fig: chi plane of mock}.

For Strong lensing, the GGSL-UNet has an extreme potential to be improved further. It can be applied to different kinds of sources by using different training samples in different survey data.
For instance, we can model AGN as background sources, as well as implement more complex Sérsic models or the Nuker model \citep{2003AJ....125.2951G} to improve the reconstruction power of the U-Net in space-based observations, or we can add more realistic contaminants in the training sample, such as spiral galaxies.

Even in the current basic version, the GGSL-UNet output is expected to represent a crucial step for strong lensing modeling if used in combination with standard MCMC or other deep learning tools such as LEMON \citep{2023MNRAS.522.5442G} to determine the corresponding strong lensing parameters, which is considered a downstream task for sophisticated analysis in the current all-sky surveys where hundreds or thousands of these events are expected to be found.

With this paper, we have started by demonstrating the effectiveness of lens image decomposition and photometry reconstruction under different noise conditions (ground-based KIDS and space-based HST). For the upcoming data from ground-based surveys, such as LSST, Wide Field Survey Telescope (WFST, \citealt{Wang2023SCPMA..6609512W}), and space-based surveys, such as {\it Euclid} \citep{2011arXiv1110.3193L}, {\it Roman} \citep{Spergel2015Roman}, and the China Space Station Telescope surveys, our tools will provide high-quality images for lens modeling and multi-band photometry reconstruction of both deflector and source.

\section*{Acknowledgements}
We acknowledge Prof. Shu Yiping for his valuable suggestions that contributed to the development of our study and Dr. Xie Linghua for her technical support. 
FZ acknowledges the support of the China Scholarship Council (grant n. 202306380249).
RL acknowledges the support from the Kunlun cluster of Sun Yat-sen University and the Starburst cluster of Purple Mountain Observatory.

\section*{Data Availability}
The code and example data are available at the following GitHub link: \url{https://github.com/Fucheng-Zhong/GGSL-UNet}.

\bibliography{references}
\bibliographystyle{aasjournal}
\appendix

\section{Anomaly cases}
\label{app: failed reconstruction of mock}
\begin{figure*}[ht]
    \includegraphics[width=1.0\textwidth]{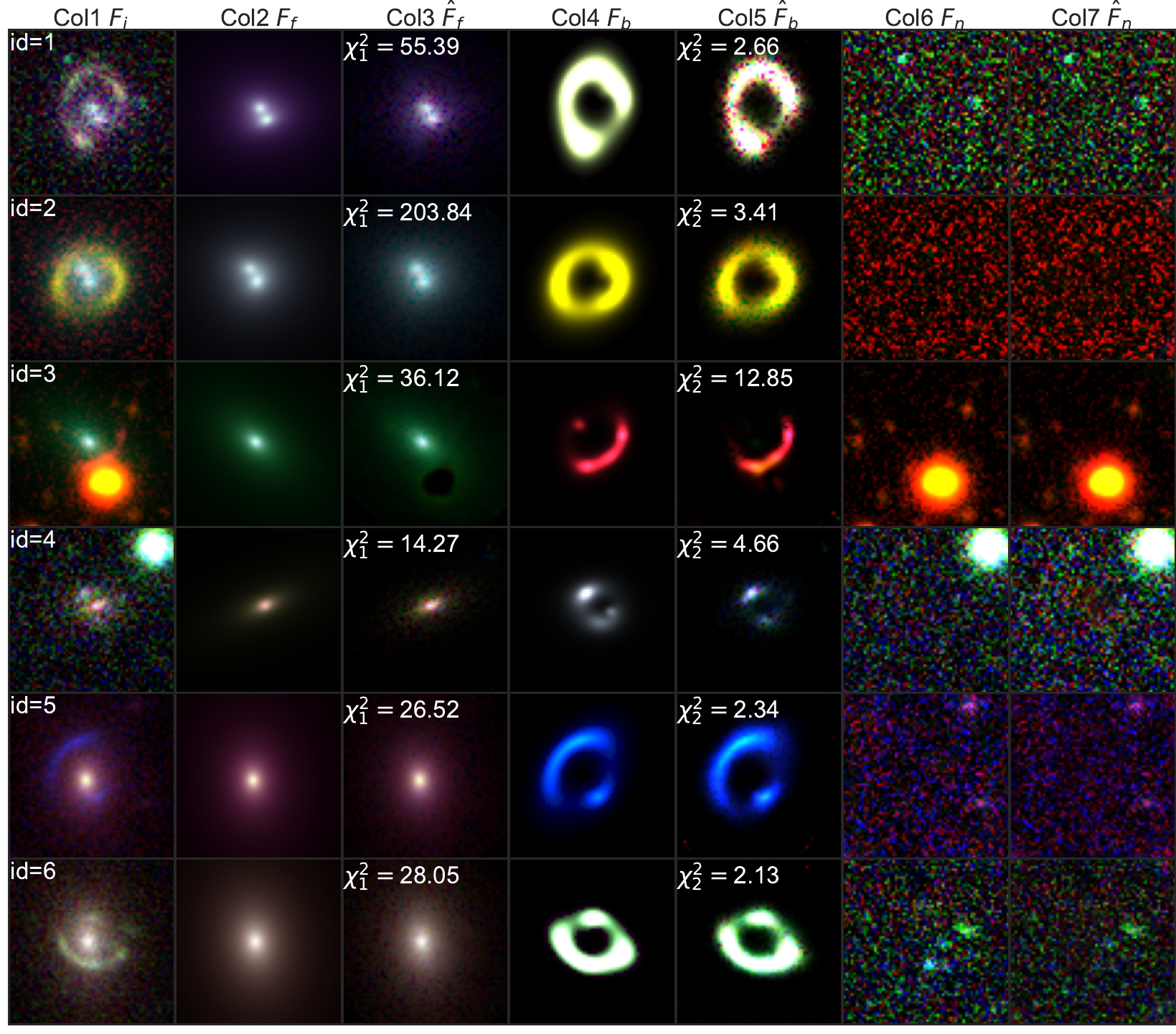}
    \caption{Anomalous reconstruction images of the mock KiDS images are shown. The labels are identical to those in Fig. \ref{fig: kids reconstructed images}. In Rows 1 and 2, the foreground contains two Sérsic models; in Rows 3 and 4, a very bright object is located very close to the background source; and in Rows 5 and 6, the background source signal is overwhelmed by noise.} 
    \label{fig: failed reconstruction images1} 
\end{figure*}

In \S\ref{sec: Evaluation}, we have used the $\chi^2$ metrics to assess the quality of the GGSL-Unet reconstruction, and we have seen that despite the majority of cases showing a $\chi_2^2<2$, $\sim20\%$ of KiDS models and $\sim 10\%$ of HST models showed a larger $\chi^2$. We stress here (as anticipated in \S\ref{sec: Method}) that, by construction, the $\chi_2^2$ value can be strongly influenced by the structure of the ``noise'' image, as strong substructures in its flux ($F_n$), which would be difficult to reconstruct by the U-Net, and then produce larger $\chi_2^2$. Hence, high $\chi_2^2$ does not generally mean bad reconstruction, as we will show below. On the other hand, there are situations where high $\chi_2^2$ values come from bad reconstructions.  
Here, we want to look into the different cases to check whether there are avenues for improvements or if there are pathological cases we need to live with. 

We start by showing some of the mock objects with large $\chi^2$ values in Fig. \ref{fig: failed reconstruction images1}, where we can distinguish the following cases: 
\begin{enumerate}
    \item Pair of foreground galaxies. Despite the fact that the GGSL-Unet can generally model these cases (see \S\ref{sec:reconstruction}), which are contemplated in the training sample, there are still cases where the reconstruction is not optimal. This is shown in rows 1 and 2 of Fig. \ref{fig: failed reconstruction images1}, where we can see the GGSL-Unet can still fit the foreground properly but with a poorer $\chi_2^2$.  
    \item Very bright ``companion'' objects (e.g., bright stars). In Rows 3 and 4 of Fig. \ref{fig: failed reconstruction images1}, the very bright objects located very close to or even overlapping with the lens strongly impact the $\chi_2^2$, although both the foreground and the background images are nicely reconstructed.
    \item Low SNR arcs. In Rows 5 and 6 of Fig. \ref{fig: failed reconstruction images1}, the background signal level is low, and most of them are overwhelmed by noise, but U-Net can still reconstruct the geometry roughly.
\end{enumerate}

\begin{figure*}
    \includegraphics[width=1.0\textwidth]{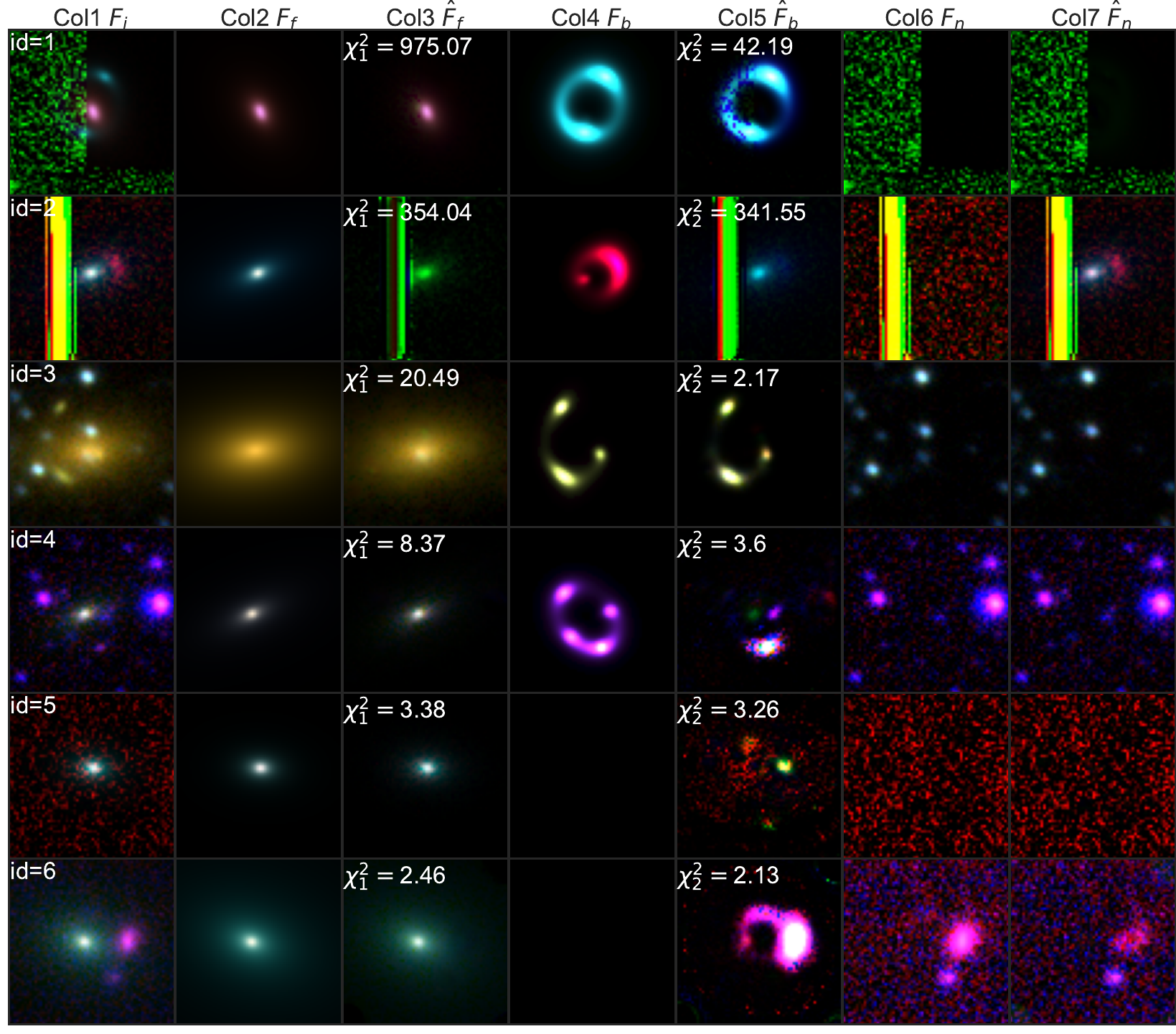}
    \caption{In Rows 1 and 2, the image is broken, with the broken part introduced by randomly selecting $F_n$ from the KiDS survey. In Rows 3 and 4, the presence of a similar extraneous source and weak background signal may cause the mis-extraction of the background. In Rows 5 and 6, some noise might be misinterpreted as a background signal.} 
    \label{fig: failed reconstruction images2} 
\end{figure*}

The ``failed'' cases are instead shown in Fig. \ref{fig: failed reconstruction images2}. Here, we can also distinguish different situations:
\begin{enumerate}
    \item Corrupted images. In Rows 1 and 2 of Fig. \ref{fig: failed reconstruction images2}, $F_n$ is randomly selected from the KiDS survey, and the edges of tiles or artificial signals make the image broken. Surprisingly, the reconstruction in Row 1 seems reasonably good (despite the large $\chi_2^2$ due to the irregular ``noise'' structure).
    \item Crowded fields. In Rows 3 and 4 of \ref{fig: failed reconstruction images2}, the weak background signal and the presence of external sources may cause the incorrect reconstruction of the background galaxy.
    \item False positive candidates. In the last two rows of Fig \ref{fig: failed reconstruction images2}, we show some no-lensing systems. Due to the absence of a background signal and heavy noise, some noise is misinterpreted as a background signal.    
\end{enumerate}

To summarise, anomalous $\chi^2$ generally imply a bad reconstruction, but, as shown in this appendix, they might also indicate a complex ``noise'' (i.e., sky background) structure. This latter case is generally true for ground-based observations, while the sharper image of space observations provides a smoother noise structure. As a strategy for practical application, one can foresee an inspection of the large $\chi^2$ cases (either using human eyeball checks or deep learning algorithm) to retain the cases with good $F_f$  and $F_b$ reconstructions.

\section{More lens reconstruction}
\label{app: More real KiDS lens} \label{sec: reconstruction of HST}
\begin{figure*}
    \includegraphics[width=0.495\textwidth]{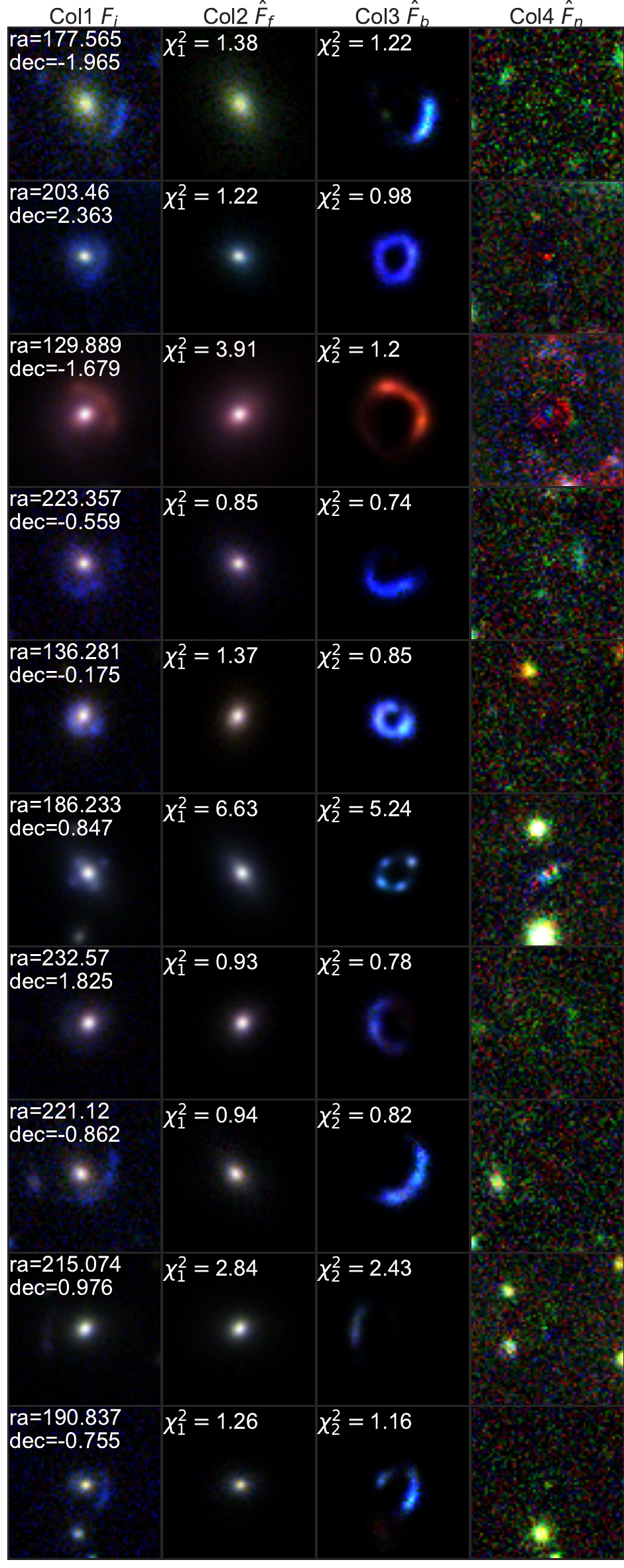}
    \includegraphics[width=0.495\textwidth]{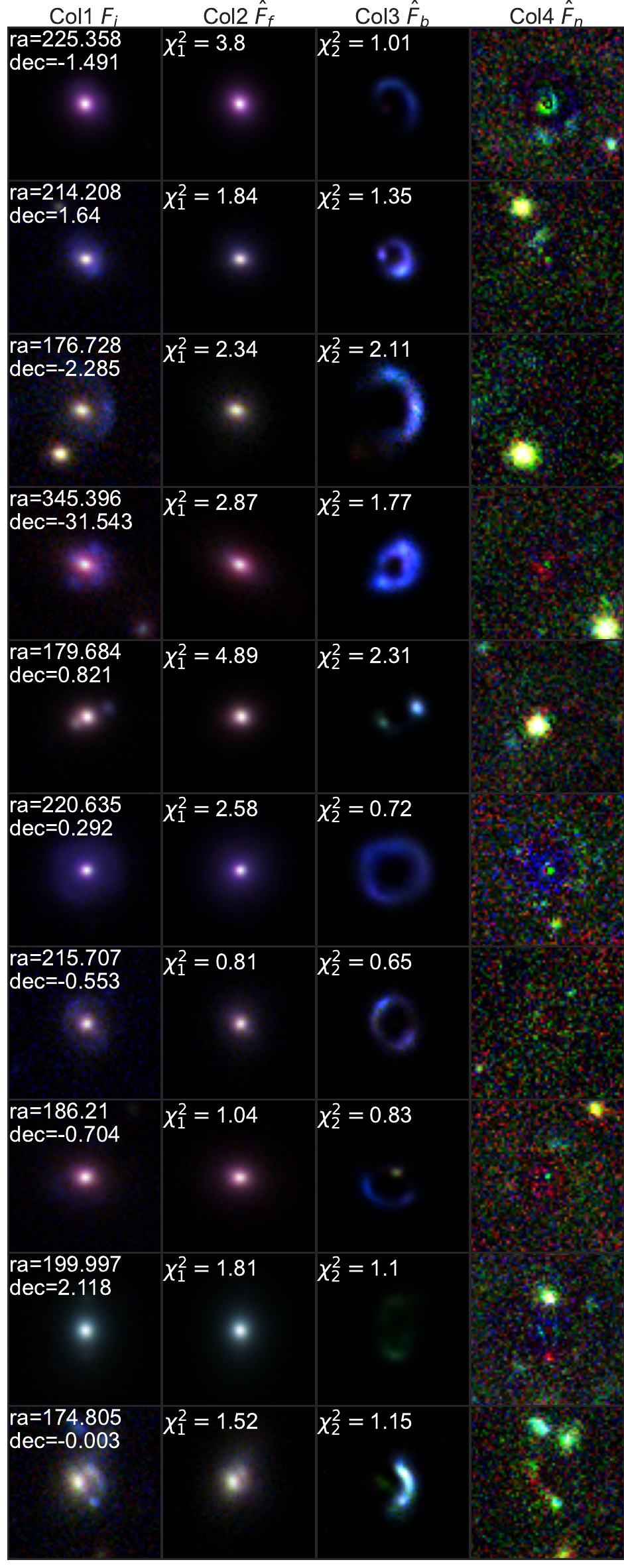}
    \caption{Modeling results for 20 real KiDS candidates. Labels are identical to those in Fig. \ref{fig: real kids reconstructed}. Some candidates might be spiral galaxies, for instance, the case of $(ra=220.635, \ dec=0.292)$.}\label{fig: more real kids images}  
\end{figure*}

We show 20 KiDS lens candidates in Fig. \ref{fig: more real kids images}. The $\chi^2$ and residuals indicate that the foreground and background components generated by U-Net are well-modeled. Most lens candidates here have low SNR, and some have extra sources besides the foreground and background components. In some cases, the foreground and background sources are very close to each other, but the U-Net can still model them very well. One issue is that the U-Net might misinterpret the arm of a spiral galaxy as a background arc. We believe this can be resolved by introducing more relevant spiral foreground training samples to the network.

\begin{figure*}
    \includegraphics[width=1\textwidth]{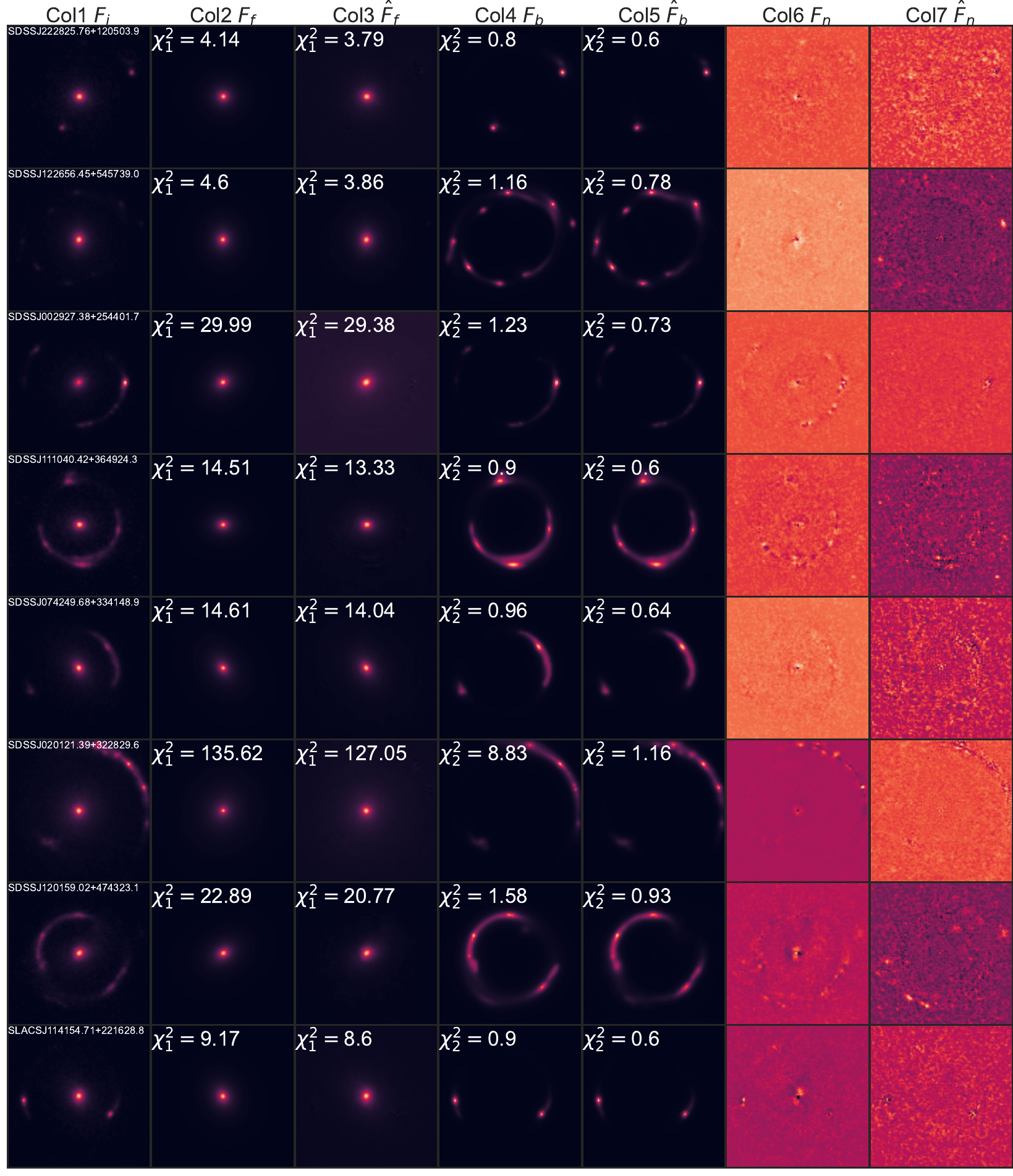}
    \caption{The BELLS GALLERY lens reconstruction results. Labels are the same as in Fig. \ref{fig: kids reconstructed images}, but now, instead of mock data, columns 2 ($F_f$) and 4 ($F_b$) show the models of BELLS GALLERY. We selected eight real BELLS GALLERY lenses to display here. The $\chi_1^2$ in column 2 and $\chi_2^2$ in column 4 are calculated by replacing $\hat{F}_f$ with $F_f$ and $\hat{F}_b$ with $F_b$ in Eq \ref{Eq: chi1} and \ref{Eq: chi2}.
    The $\chi^2$ and residual $\hat F_n$ from the U-Net show a reconstruction quality comparable to the residual $F_n$ from BELLS GALLERY.}
    \label{fig: real HST images}
\end{figure*}

We also show the 8 BELLS GALLERY images (only $r$-band) reconstructed by the U-Net in Fig. \ref{fig: real HST images}. 
Differently from the KiDS cases, here we can compare the GGSL-Unet reconstruction with the ray-tracing model by \citet{Shu2016ApJ...833..264S}. Columns 2 and 4 in the figure represent models of standard classical ray-tracing methods. We include them here to compare the deep learning and traditional MCMC-based classical methods.
By comparing the $\chi_2^2$, we can claim that the U-Net modeling achieves at least the same level of accuracy as the classical method, and in some cases, it performs even better.

\section{Downstream task: Photometric redshift}
\label{sec: Downstream task}

In \S\ref{sec: Evaluation}, we have demonstrated that 1) the U-Net can effectively denoise and ``decouple'' the foreground galaxy and background components from the input lens image; 2) the U-Net can also reconstruct the intrinsic photometry of the strong lensing components, although, for the ground-based observations, there are systematic offsets between estimated magnitudes and true values.

To illustrate a practical application of the GGSL-Unet output, we have estimated the photometric redshift as a downstream task. We calculate the magnitude of different bands from reconstructed fore/background images and obtain the photometric redshift for both components.

\begin{figure} 
        \centering
        \includegraphics[width=0.45\textwidth]{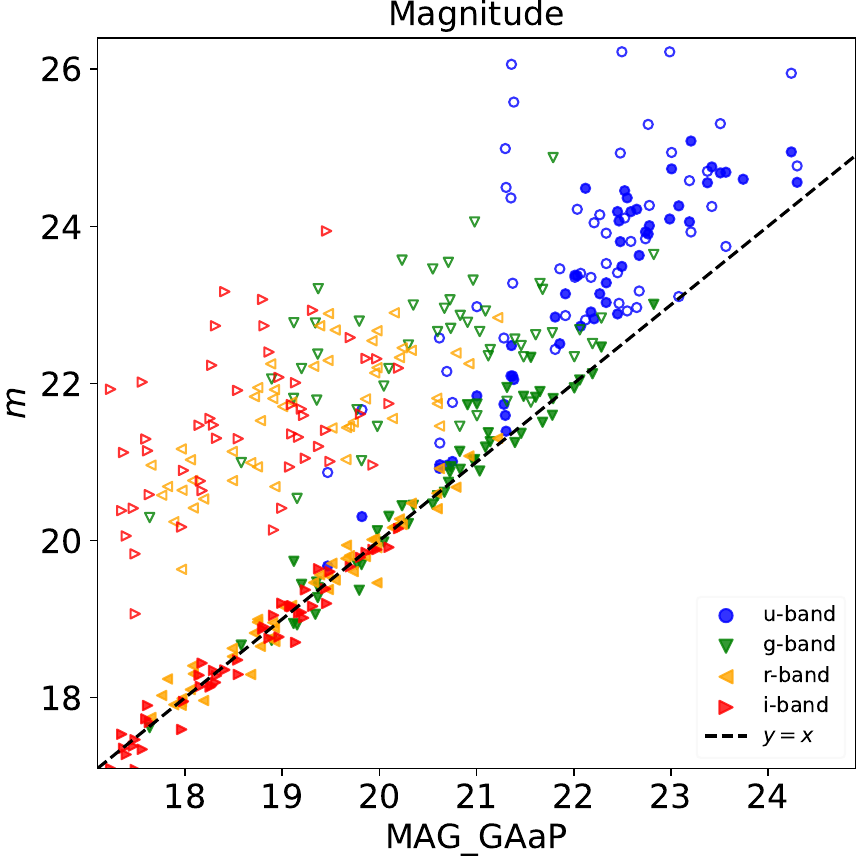}
    \caption{The optical band magnitudes calculated from the U-Net flux are compared to the magnitudes provided by the KiDS catalog for 50 lens candidates. Those magnitudes include the extinction corrections. The empty circles label the background source magnitudes, while the filled circles represent the foreground. The KiDS magnitudes are calculated using the optimal minimum Gaussian Aperture and Photometry (GAaP) flux for each source, and extinction corrections are also included.}    
    \label{fig: magnitude} 
\end{figure}

\begin{figure} 
        \centering
        \includegraphics[width=0.45\textwidth]{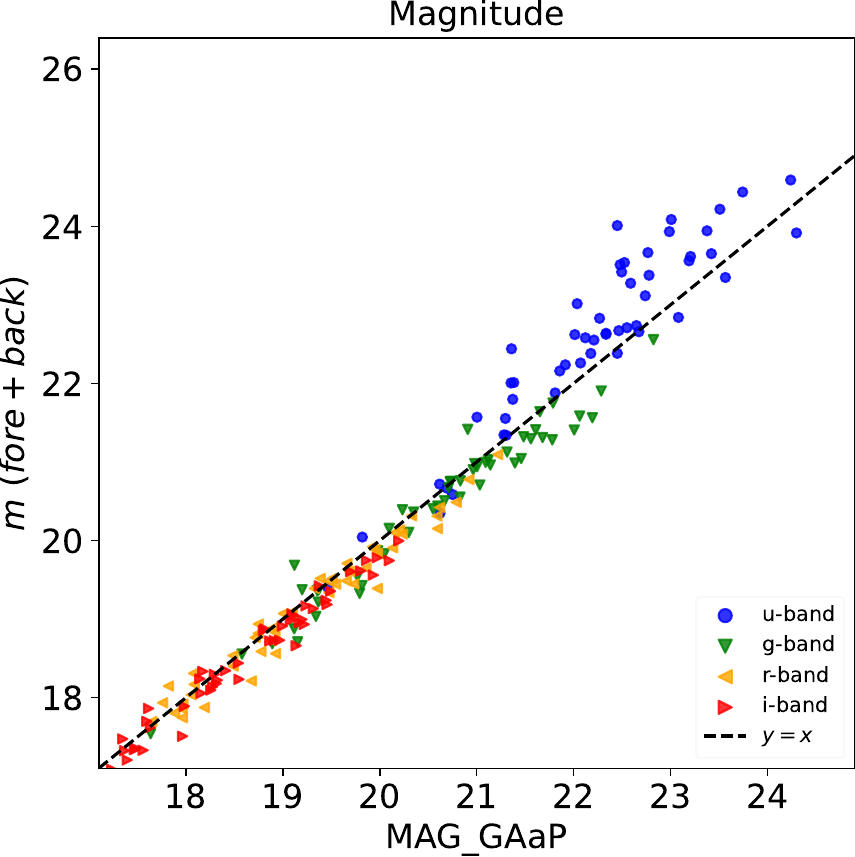}
    \caption{The optical band magnitudes calculated from the U-Net flux $\hat F_f + \hat F_b$ are compared to the magnitudes provided by the KiDS catalog for 50 lens candidates. The labels are identical to those in Fig. \ref{fig: magnitude}.}    
    \label{fig: total magnitude} 
\end{figure}

The aperture magnitude of the foreground and background sources are approximately calculated following Eq. \ref{eq: magnitude}.
In Fig. \ref{fig: magnitude}, we plot the foreground and background magnitudes, $m_f$ and $m_b$, which are calculated using the reconstructed optical band flux $\hat{F}_f$ and $\hat{F}_b$, versus the KiDS catalog magnitude, MAG\_GAaP in Table \ref{table:51_KiDS_lens}. Here, we have corrected $m_f$ and $m_b$ for extinction, as has already been done for the KiDS catalog magnitudes (MAG\_GAaP, see \citealt{Kuijken2019A&A...625A...2K}). 
The $m_b$ is much larger than $m_f$, as expected, since most of the lens candidates in the dataset have lower flux compared to the foreground galaxy.
Considering that the foreground flux is difficult to distinguish from the background and noise when the flux is low, we have additionally applied a correction for the foreground, using the flux inside the 95\% flux contour of $\hat{F}_f$. One would expect $m_f$ to approximate MAG\_GAaP, but it is slightly larger in the $u$ band. 
As MAG\_GAaP is derived from galaxy shapes and position angles, the GAaP method may not be able to distinguish between the lens and arcs in low SNR images, such as those in the $u$-band. Consequently, MAG\_GAaP values for $u$-band images might include contributions from both the lens and the background galaxy.
\Zhong{Because the flux $\hat F_f$ only includes the flux of the foreground, we calculate the total flux of the foreground, $\hat F_f$, and background, $\hat F_b$. The magnitude is then calculated using Eq \ref{eq: magnitude}, and the results are shown in Fig. \ref{fig: total magnitude}. The figure shows that the $u$-band matches better when considering both foreground and background flux, but the $u$-band magnitudes are still slightly overestimated in this case. Nonetheless, it demonstrates that the background image contributes a significant portion of the $u$-band flux. This is consistent with the physical scenario of the lens where the foreground galaxy is redder while the background is bluer.}

From the figure, we can see that for the foreground, the $g$, $r$, and $i$ magnitudes are well reproduced by the U-Net output. However, the deviation in the $u$-band magnitude is considerably larger. This is because the $u$-band flux is much smaller than the other redder bands, making the noise and background contribute much higher than in other bands, hence causing larger relative error. 

\begin{figure} 
        \centering
        \includegraphics[width=0.45\textwidth]{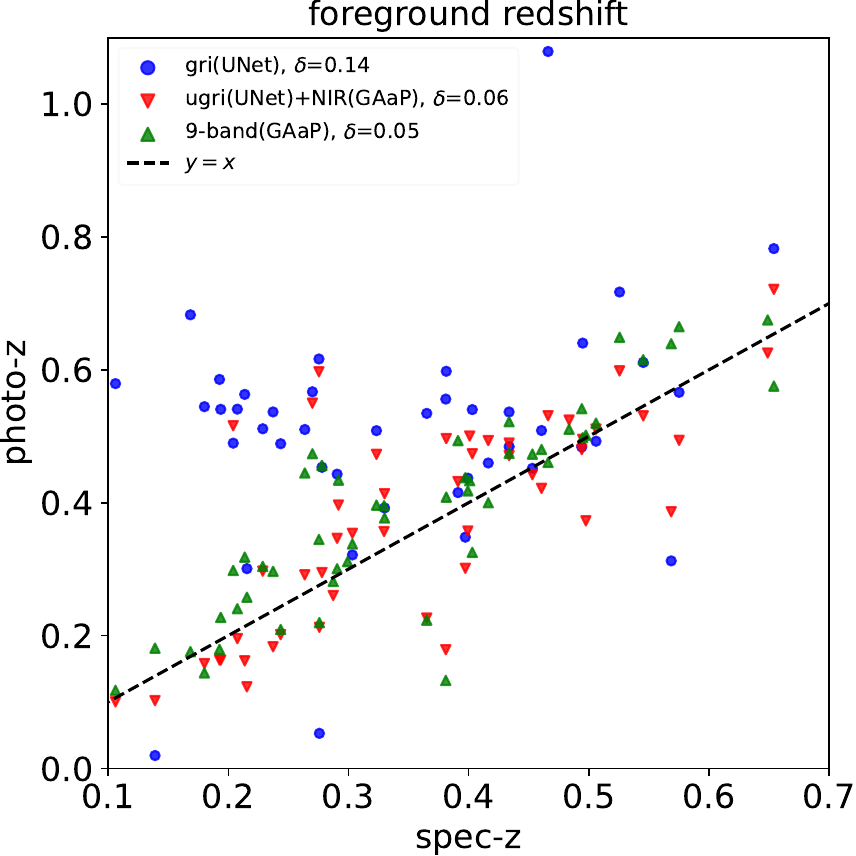}
    \caption{The photometric redshift versus the spectroscopic redshift for the foreground is shown. The mean relative redshift error is calculated as follows: $\delta = |\text{photo-z} - \text{spec-z}|/(1+\text{spec-z})$. The photo-z is measured using magnitudes in three different ways: (1) using U-Net $gri$ band magnitudes (blue points), (2) using U-Net optical magnitudes combined with Near Infrared GAaP magnitudes (for bands other than the optical bands, red points), and (3) using 9-band GAaP magnitudes (green points).}    
    \label{fig: photo-z} 
\end{figure}

Due to the large uncertainties in the $u$-band magnitude, we predict the foreground redshift using the remaining bands, $g$, $r$, and $i$. We match the corresponding targets with the spectroscopic measurements and use the spectroscopic redshift as the ground truth. In Fig. \ref{fig: photo-z}, we demonstrate the foreground photometric redshift estimated by $\it LePHARE$ versus the spectroscopic redshift. $\it LePHARE$ photo-$z$have been obtained using stellar templates from \cite{Bruzual2003} with a \cite{Chabrier2003} initial mass function (IMF) and an exponentially decaying star formation history. The photometric redshift is calculated using magnitudes either provided by the KiDS catalog or derived from Eq. \ref{eq: magnitude} using U-Net outputs.
As mentioned for the foreground galaxy, we have estimated the photometric redshift using only the magnitudes of three bands ($gri$). However, for comparison, we have also estimated the photometric redshift using 9-band magnitudes, adding the near-infrared (NIR) magnitudes from $ZYHJK_s$ from the KiDS/VIKING catalog (see again \citealt{Kuijken2019A&A...625A...2K}). We also use other data combinations, like the $ugri$ magnitudes from U-Net combined with the $ZYJHK_S$ magnitudes from the KiDS.
The different photo-$z$ estimates of the foreground galaxy from the photometric combinations above are shown in Fig. \ref{fig: photo-z}.
Here, we can clearly see that the photo-$z$ based on optical band only is very poorly estimated (scatter $\delta=0.14$) as compared to the full 9-band photometry for which we have found a scatter $\delta=0.05$ and all estimates aligned along the 1-to-1 relation. However, we can see that the GGSL-Unet photometry also allowed us to obtain similar accuracy and precision ($\delta=0.06$) of the 9-band KiDS photometry, showing a substantial consistency for the foreground galaxy and the reliability of our photometry reconstruction in the optical bands, although we have assessed the presence of systematics of 0.03 mag (see \S\ref{sec:photometry_reconstr}).

To make the U-Net model fully multi-band (from optical to NIR) and possibly achieve higher accuracy in redshift estimation, we need to incorporate additional infrared bands into the U-Net training. This is one of the next steps of our project for upcoming papers.

\begin{figure} 
        \centering
        \includegraphics[width=0.45\textwidth]{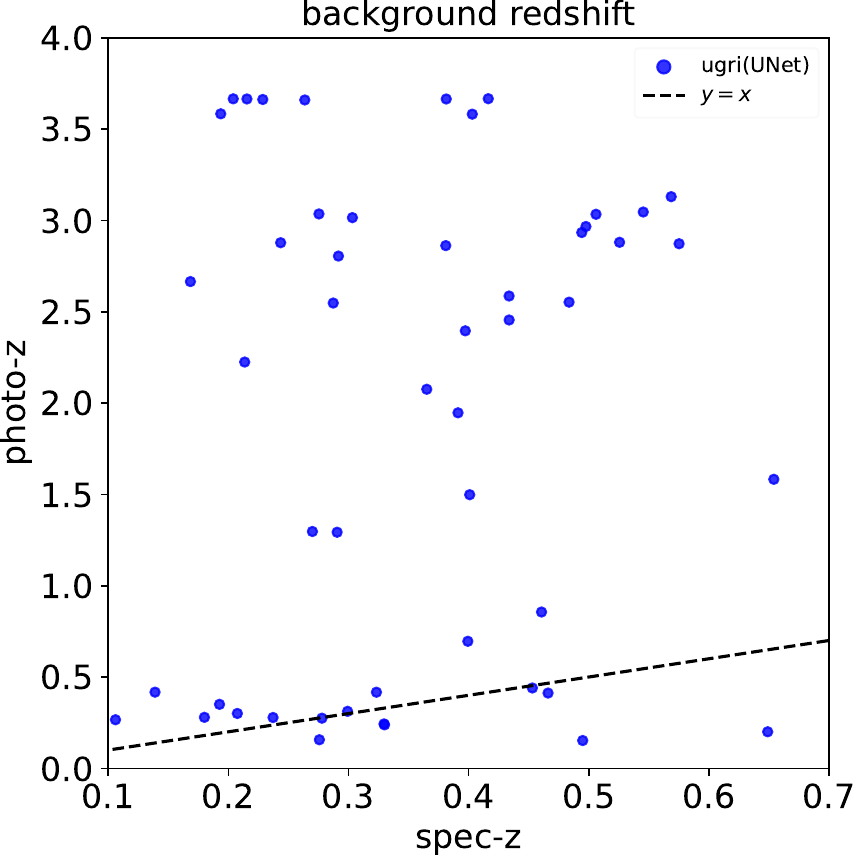}
    \caption{The estimated photometric redshift (via LePHARE) of the background versus the spectroscopic redshift of the foreground is shown. Comparing the background and lens redshifts provides a possible way to distinguish non-lensing contamination among the lens candidates. The photo-z is measured using magnitudes from four optical bands, which are directly calculated from the U-Net output background flux.}
    \label{fig: back photo-z} 
\end{figure}

This important implementation will be fundamental for accurate photo-z of the background sources. To derive the integrated photometry of the background galaxies in the optica+NIR, we also need the GGSL-Unet to perform in NIR images. Currently, we have tried to assess the accuracy of the photo-$z$ based only on the U-Net optical photometry. This is shown in Fig. \ref{fig: back photo-z}, where the background redshift estimated using four optical bands ($ugri$) is plotted against the spectroscopic redshift of the foreground as a sanity check. \Zhong{Here, we have also considered the $u$-band in the photo-$z$ determination, despite the systematics discussed above, but not as severe for the background source as the foreground.
} 

The figure shows that the majority of the photo-$z$ of the background source is larger than the spectroscopic redshift of the foreground system, hence suggesting that this is compatible with a lensing system. 
We want to stress that if we can determine the background redshift, we can further distinguish whether the candidates are true lenses or just spiral/diffuse galaxies because the background redshift should be larger than the foreground spec-z or photo-z. In contrast, spiral or diffuse galaxies have similar foreground and background redshifts. On the other hand, for $\sim15$ systems (i.e., $\sim 30\%$ of the full catalog), the photo-$z$ of the source is similar if not smaller than the spec-$z$ of the deflector, suggesting that these photo-$z$ are likely to be wrong.

\label{lastpage}
\end{document}